\documentclass[pra,showpacs,preprintnumbers,amsmath,amssymb,tightenlines,twocolumn,superscriptaddress]{revtex4}
\usepackage{graphicx}
\usepackage{dcolumn}
\usepackage{bm}
\usepackage{epsfig}
\usepackage{enumerate}
\usepackage[utf8]{inputenc}
\usepackage{float}
\usepackage{siunitx}
\newcommand{\ssection}[1]{{\noi  \it #1:}}

\newcommand{\ket}[2][]{\mathinner{\lvert#2\rangle}_{\hspace{-0.1em}#1}}

\newcommand{\braket}[2]{\mathinner{\langle{#1\lvert#2}\rangle}}
\newcommand{\ketbra}[2]{\mathinner{\lvert#1\rangle\langle #2\rvert}}
\newcommand{\matrixelem}[3]{\mathinner{\langle{ #1 \lvert #2 \lvert #3} \rangle}}

\newcommand{\sub}[2]{{#1}_{\mbox{\!\! \scriptsize #2}}}

\newcommand{\bv}[1]{\mathbf{ #1 }}

\def\noi{\noindent}
\def\beq{\begin{equation}}
\def\eeq{\end{equation}}

\def\CR{\nonumber\\[0.15cm]}

\newcommand{\fref}[1]{Fig.~\ref{#1}}
\newcommand{\frefp}[2]{Fig.~\ref{#1}~(#2)}

\newcommand{\eref}[1]{Eq.~(\ref{#1})}

\newcommand{\sref}[1]{section~\ref{#1}}

\newcommand{\cref}[1]{chapter~\ref{#1}}

\newcommand{\Cref}[1]{Chapter~\ref{#1}}

\newcommand{\bref}[1]{(\ref{#1})}

\usepackage{bbm}
\usepackage{MnSymbol}
\usepackage{afterpage}
\makeatother
\newcommand{\astate}{a} 
\newcommand{\bstate}{b} 

\binoppenalty=\maxdimen

\begin{document}

\title{On-chip quantum tomography of mechanical nano-scale oscillators with guided Rydberg atoms}
\author{A.~Sanz-Mora}  \affiliation{Max Planck Institute for the Physics of Complex Systems, N\"othnitzer Strasse 38, 01187 Dresden, Germany} 
\author{S.~W\"uster} 
\affiliation{Max Planck Institute for the Physics of Complex Systems, N\"othnitzer Strasse 38, 01187 Dresden, Germany}
\affiliation{Department of Physics, Bilkent University, 06800 {\c C}ankaya, Ankara, Turkey}
\affiliation{Department of Physics, Indian Institute of Science Education and Research, Bhopal, Madhya Pradesh 462 023, India}
\author{J.-M.~Rost} \affiliation{Max Planck Institute for the Physics of Complex Systems, N\"othnitzer Strasse 38, 01187 Dresden, Germany}
\begin{abstract}
Nano-mechanical oscillators as well as Rydberg-atomic waveguides hosted on micro-fabricated chip surfaces hold promise to become pillars of future quantum technologies. In a hybrid platform with both, we show that beams of Rydberg atoms in waveguides can quantum-coherently interrogate and manipulate nanomechanical elements, allowing full quantum state tomography. Central to the tomography are quantum non-demolition measurements using the Rydberg atoms as probes. Quantum coherent displacement of the oscillator is also made possible, by driving the atoms with external fields while they interact with the oscillator. We numerically demonstrate the feasibility of this fully integrated on-chip control and read-out suite for quantum nano-mechanics, taking into account noise and error sources.
\end{abstract}
\pacs{
  32.80.Ee,     
  37.10.Gh,     
  07.10.Cm,    
  03.65.Wj  
}

\maketitle
\ssection{Introduction}
%
Quantum optomechanics was originally developed in the context of gravitational wave detection \cite{abbot_gravwave_PRL}. Subsequently, it took up the challenge to cool nanoscale quantum systems all the way to their ground state \cite{chan:groundstatecooling}, and more generally to gain the level of control over them that we are used to have over quantum optical systems \cite{aspelmeyer:review,kippenberg:review,poot:mechquantumsyst}, e.g, through coupling to non-classical light \cite{Palomaki_osc_mw_entanglement_science,Riedinger_nonclass_mechlight_nature}.
Experiments on the quantum non-demolition (QND) read-out of the phonon state of a nanomechanical oscillator or its state-tomography have  only  begun recently \cite{Lecocq_QNDnonclassical_PRX, Lei_2016_Nondemolition} and most existing proposals  interface  the oscillator with a cavity \cite{vanner:statereconstr,Woolley_2010_quantum,Gangat_2011_number,Ludwig_2012_Quantum,Yin_2013_quantum}.
\begin{figure}[htb]
\begin{centering}
\includegraphics[width=\columnwidth, trim={0 1.2cm 0 0}]{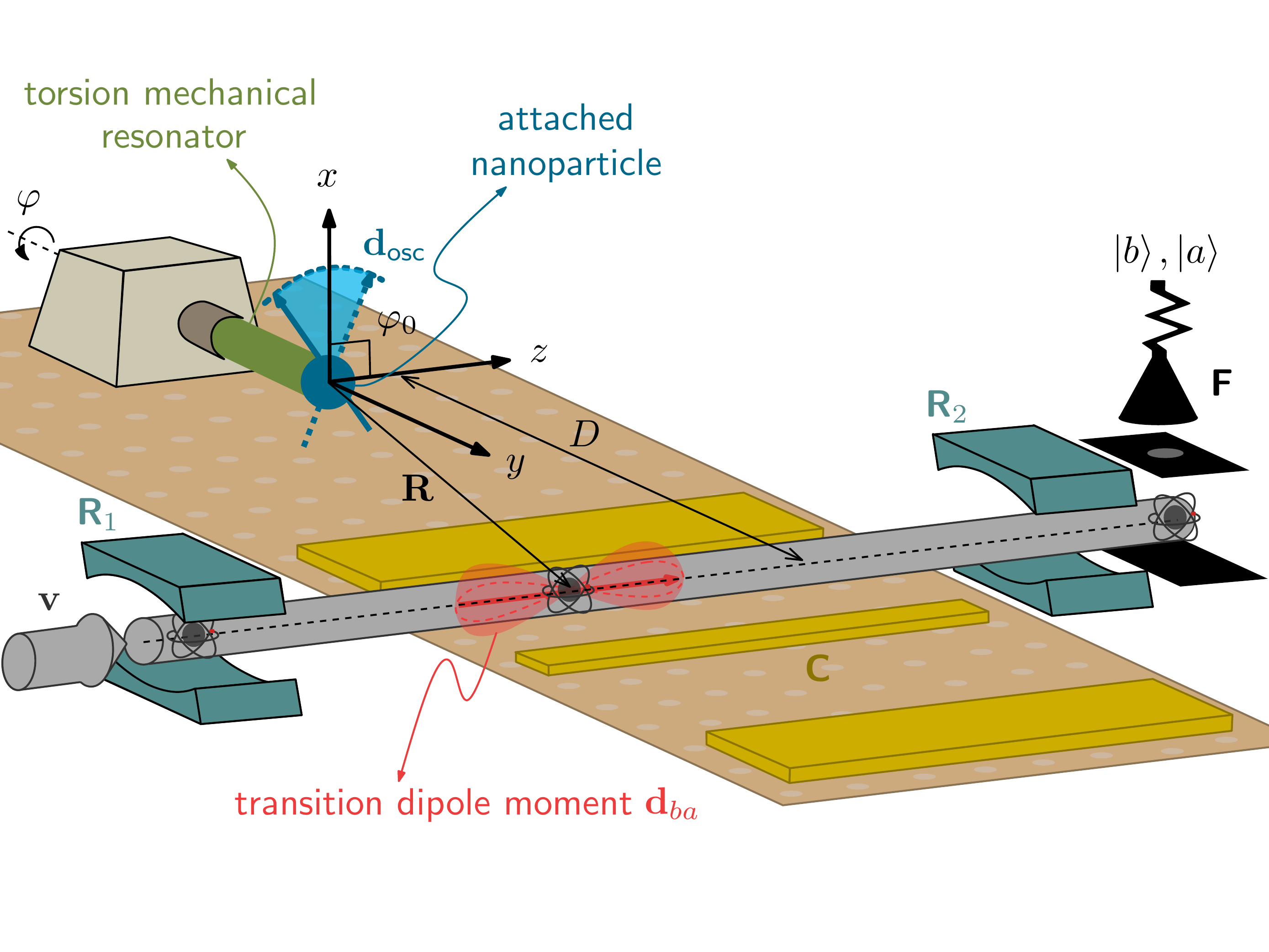}
\par\end{centering}
\begin{centering}
  \caption{\label{fig:Geometric_configuration}
    (Color online) schematic of the coupled Rydberg atomic wave-guide (grey) and torsion mechanical oscillator (green). The beam passes the oscillator with  impact parameter $D$ and velocity $\mathbf{v}$. Oscillations of the torsion angle $\varphi$ modulate the interaction between the permanent electric moment $\mathbf{d}_{\text{osc}}$ of a ferroelectric load (blue) and the transition dipole moment $\mathbf{d}_{\bstate \astate}$ involving atomic Rydberg states $\ket{\bstate}$ and $\ket{\astate}$.  While a train of atoms interacts one-by-one with the oscillator, the states incur a phase-shift dependent on the state of the oscillator. This shift is interferometrically read out using microwave $\pi/2$-pulses in regions $\text{R}_{1/2}$ and state detection in $\text{F}$. Quantum coherent manipulation of the oscillator for quantum tomography uses additional external driving of the atoms through a coplanar microwave waveguide in region $\mathrm{C}$.
}
\par\end{centering}
\end{figure}

For torsional oscillators \cite{kim_quantlim_torque_NComm,kim_torsional_optmech}, we develop in the following a scheme without direct cavity interfacing, allowing for integration of mechanical- and measurement elements into the same nano-fabricated substrate  using Rydberg atoms. With their long life times
and strong long-range interactions \cite{book:gallagher}, they mingle naturally with the time- and spatial scales of optomechanics. Furthermore, with accessible  atomic transition frequencies spanning orders of magnitude when varying the principal quantum number $\nu$, Rydberg atoms as an interface partner promise  to preserve the wide range of oscillation frequencies which can be generated from nano-mechanical elements
\cite{poot:mechquantumsyst,aspelmeyer:review,kippenberg:review}.

Recent advances in manipulation and control of Rydberg atoms through on-chip waveguides \cite{Hogan_2011_Electrode, Lancuba_2013_Rydberg, Lancuba_2014_line} as well as in retaining atomic coherence closer to chip surfaces \cite{Carter_2012_field, Carter_2013_manipulation,Thiele_2014_Rydberg,Avigliano_2014_coherence} render  Rydberg on-chip integration promising and realistic by matching   Rydberg atom interaction ranges with  the spatial $\mu$m scales of the chip geometry.

To achieve a full quantum tomography of the torsional nano-oscillator, the Rydberg atoms have to fullfill a twofold role: Firstly, the atomic Rydberg beam in the waveguide passing by the oscillator acts as probe for the oscillator state \cite{keil:atomchip:review,Hogan_2016_Stark}:
Controlled electro-static interactions between the oscillator and the atoms cause a phonon-number dependent phase shift, to be read out interferometrically  \cite{BrHaLe90_976,brune:processtomog,BrHaRa92_5193,GlKuGu07_297,deleglise:reconstruction}. 
Secondly, we can coherently displace the nano-mechanical oscillator
 by externally driving the Rydberg atom while it is passing by the oscillator in order to scan the oscillator state in a controlled fashion. In this step the Rydberg atom acts as a mediator for quantum control of the oscillator. 
  
So far, destructive Wigner tomography \cite{singh:wignerntomog} has been proposed, was well as the use of 
classical oscillators for atomic quantum state control \cite{singh:cantilevermol}, which could also be achieved through interaction of nano-mechanical elements with atoms or molecules \cite{bariani:controlwsingleatom,stevenson:chargedosci:rydberg}.
 
We extend these works by transferring Rydberg-atom based QND detection developed in the context of cavity-QED \cite{BrHaLe90_976,brune:processtomog,BrHaRa92_5193,GlKuGu07_297,deleglise:reconstruction} to the realm of quantum nano-mechanics, and integrate all these functional elements into a versatile on-chip Rydberg atomic probe technique without the need of a cavity. 
   
\ssection{Hybrid setup of nano-oscillator and Rydberg waveguide}
%
While  our scheme is  quite general, we nevertheless will consider a carbon nanotube (CNT) as torsional oscillator to be specific. Interfacing the Rydberg waveguide, we will explicitly take into account the expected dominant decoherence sources. The CNT is clamped to a chip-surface and equipped with a weight at one end, as, e.g., in \cite{MePaRo05_1539, Ganzhorn_2013_spin}, which allows for tuning  the oscillation frequency. The weight will be a ferroelectric nanoparticle with a permanent electric dipole moment $\mathbf{d}_{\text{osc}}$~\cite{Basun_2011_moment}, providing a simple and adjustable 
interaction between oscillator and Rydberg atoms, independent of surface and material properties. The Rydberg atoms,  confined to an atomic waveguide \cite{Hogan_2012_Rydberg,keil:atomchip:review}, pass the oscillator with impact parameter $D$ as shown in \fref{fig:Geometric_configuration}. 

First, the atoms in the guide are excited to a Rydberg state $\ket{\astate}$ (not shown in \fref{fig:Geometric_configuration}). In the region $\text{R}_1$ of \fref{fig:Geometric_configuration} the waveguide passes a micro-wave cavity, which generates a Rabi $\pi/2$ pulse tuned to  the transition between $\ket{\astate}$ and a second Rydberg state $\ket{\bstate}$. As a consequence,
 atoms in the  state $[\ket{\astate}+\ket{\bstate}]/\sqrt{2}$ are produced.  

Parameters are chosen such that the interactions with the oscillator (in the coupling region $\text{C}$)
 effect only a relative phase shift between  $\ket{\astate}$ and $\ket{\bstate}$, to be inferred 
 from detection of the Rydberg state at $\text{F}$ after 
 a second $\pi/2$ pulse in region $\text{R}_2$. If the oscillator is in a Fock state, such a Ramsey interferometric measurement leaves the oscillator state unchanged, thus furnishing a QND-measurement. For more general oscillator states, a series of these measurements will gradually collapse the state towards a phonon number (Fock) state \cite{Guerlin_2007_field}. Repeating such series multiple times eventually reveals the entire phonon-number distribution. 
 
 Full quantum-state tomography  requires knowledge of the phases between different number states, which can be obtained after quantum coherently displacing the oscillator prior to the phonon-number distribution measurement. To obtain a well defined displacement, we propose to externally drive Rydberg atoms while they pass through the strong interaction region $\text{C}$ as discussed below. For a well defined coupling the driving should target only the Rydberg atoms and not directly the oscillator by using well localized evanescent fields from a coplanar microwave guide \cite{Beck_2016_coplanar,Thiele_2015_electric} or a three-photon off-resonant Raman transition \cite{Melo_2015_shift}.
 
\ssection{Model}
%
We formalize our scheme with the Hamiltonian  
\begin{align}
\hat{H}&=\hat{H}_{\text{osc}}+\hat{H}_{\text{at}}+\hat{H}_{\text{int}} + \hat{H}_{\text{coup}}
\label{eq:totalhamil}
\end{align}
to demonstrate quantitatively the feasibility of this protocol.
The oscillator with frequency $\sub{\omega}{osc}$ is described by $\hat{H}_{\text{osc}}=\hbar \sub{\omega}{osc}\hat{c}^{\dagger}\hat{c}$, with corresponding oscillator states $\ket{n}$ and ladder operators $\hat{c}$, $\hat{c}^{\dagger}$. The Hamiltonian for the internal state of a single atom is $\hat{H}_{\text{at}} =\hbar\omega_{\bstate\astate}\hat{\sigma}_{\bstate\bstate}$, where $\hat{\sigma}_{\mu'\mu}=\ketbra{\mu'}{\mu}$ denotes the atomic transition operator between levels $\ket{\mu'}$ and $\ket{\mu}$, and $\omega_{\mu'\mu}$ the corresponding Bohr frequency. Motion of the atoms in the waveguide is treated classically as described below.
The atom-oscillator coupling $\hat{H}_{\text{int}}$ is due to electric dipole-dipole interactions between the transition dipole of the atom and the permanent dipole of the nano-particle attached to the oscillator. By choosing the atomic transition dipole $\mathbf{d}_{\bstate\astate} = \matrixelem \bstate{\hat{\mathbf{d}}_{\text{at}}}\astate$ ($\hat{\mathbf{d}}_{\text{at}}$ is the atomic dipole operator) along the $z$-axis and for an atom precisely at the centre of the waveguide, we find \cite{sup:inf}
\begin{align}
\hat{H}_{\text{int}} &=-\hbar{\cal K}(\mathbf{R})[\hat{c}^{\dagger}+\hat{c}][\hat{\sigma}_{\bstate\astate}+\hat{\sigma}_{\astate\bstate}].
\label{eq:rabi_coupling}
\end{align}
Here the interaction strength is ${\cal K}(\mathbf{R}) = {\cal K}_0 f(\mathbf{R})$, where ${\cal K}_0 = V_0/\sqrt{2\hbar\omega_{\text{osc}}I}$, with $V_0 = \Vert\bv{d}_{\bstate \astate}\Vert \Vert \bv{d}_{\text{osc}}\Vert/[4\pi\varepsilon_{0}D^{3}]$, and $I$ the moment of inertia of the oscillator (CNT and nanoparticle). The vector $\bv{R} = \bv{R}(t)$, with $R = \Vert\mathbf{R}\Vert$, points from the center of the nanoparticle in equilibrium (origin of our coordinate system) to the atom in the waveguide, as shown in \fref{fig:Geometric_configuration}. Then the interaction amplitude becomes $f(\mathbf{R}) = [D/R]^3[1 - 3Z^2/R^2]$. We assumed small excursions of the oscillator from an equilibrium torsional angle $\varphi_{0}=\pi/2$. 

Finally, $\hat{H}_{\text{coup}} = \hbar\mathnormal{\Omega}(t) [\hat{\sigma}_{\bstate\astate}+\hat{\sigma}_{\astate\bstate}]/2$ represents the controllable inter-state resonant coupling with Rabi frequency $\mathnormal{\Omega}(t)$ in dipole and rotating wave approximations. This term is due to a microwave field in regions R$_{1}$, R$_{2}$ and possibly  $\text{C}$.

\ssection{Ramsey measurement of phonon number}
%
We consider a scenario where the atomic transition frequency $\omega_{\bstate\astate}$ is much closer to resonance with the oscillator $\omega_{\text{osc}}$ than any other transition frequency. This justifies taking into account  atomic states $\ket{a},\ket{b}$ only. There are two advantages in choosing these as Rydberg states: (i) 
For a wide range of mechanical oscillation frequencies $1$ MHz $<\omega_{\text{osc}}<10$ GHz, some near resonant atomic transitions can be found with $\omega_{\bstate\astate}=\omega_{\text{osc}}+\delta$ and atom-oscillator detuning $\delta$ much smaller than energy gaps to any other atomic states. (ii) The large Rydberg transition dipoles $\mathbf{d}_{\bstate \astate}$ provide strong coupling ${\cal K}_0$ between atom and oscillator without too close proximity. 
%
%
\begin{figure}[htb]
  \centering
    \includegraphics[width=0.99\columnwidth]{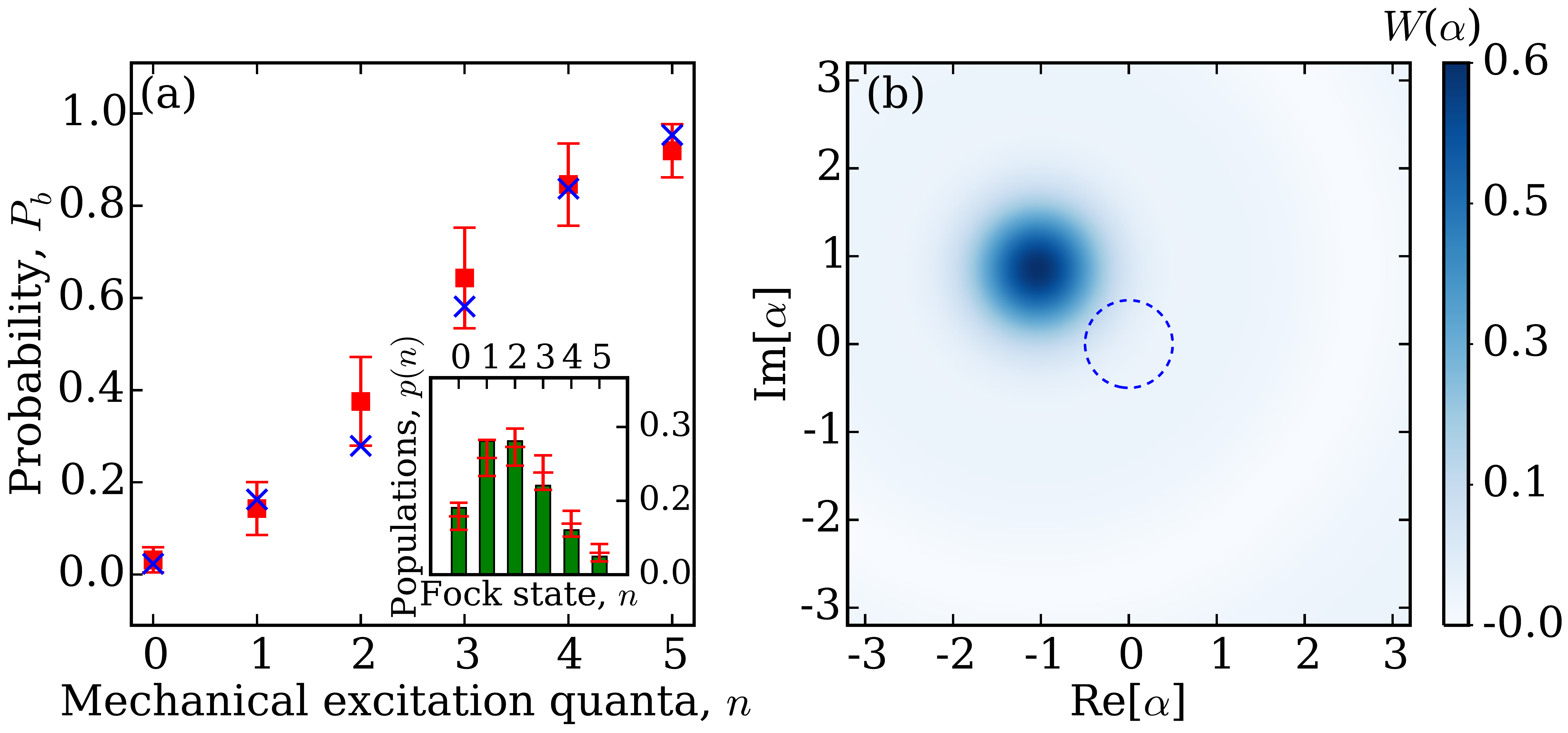}
\caption{\label{fig:phaseshifts_displacement}
(Color online) (a) Phonon-state dependent probability $P_b$ for the probe atom to arrive in state $\ket{b}$. (Blue $\times$) ideal result for an atom at waveguide centre, (red) average and standard deviation with finite width atomic beam. Both data sets include decoherence. The inset shows the inferred phonon distribution for a coherent oscillator state $\ket{\alpha}$ with coherent amplitude $\alpha=\sqrt{2}$.
(b) Wigner function of mechanical oscillator state. Before displacement we take the ground state (contour), after displacement operation $\hat{{\cal D}}(\alpha_N(\tau))$ we obtain a coherent state (color). We used $\alpha_N(\tau)/|\alpha_N(\tau)|=-[1- \mathrm{i}]/\sqrt{2}$, with $|\alpha_N(\tau)|\simeq 1.35$, $\mathnormal{\Omega}/(2 \pi) = \sqrt{2}/2 \, \mathrm{MHz}$, see text.
}
\end{figure}
Nonetheless, we can achieve a situation where the atom and oscillator are far off-resonant with respect to the coupling strength ${\cal{K}}(\bv{R})$. Then creation and destruction of phonons through \bref{eq:rabi_coupling} are suppressed. Thus, while the compound system of atom plus oscillator adiabatically follows its interacting eigenstates as the atom passes the oscillator, the interaction only causes a  phase shift $\Phi^{(n)}[\delta]$ between the superimposed states $\ket{\astate}$, $\ket{\bstate}$ depending on the detuning $\delta$ and the phonon number $n$. After the second Rabi $\pi/2$ pulse at $\text{R}_2$, this phase shift  affects the results of destructive state measurement in $\text{F}$. Ultimately, after a succession of such measurements we may infer $\cos(\Phi^{(n)}[\delta])$. Further details can be found in \cite{sup:inf} and \cite{Haroche_2006_the}.

We now proceed to simulate measurements, taking into account de-coherence sources to explore the practical limitations arising through the vicinity of a micro-chip surface and the Rydberg atom waveguide.  We work in a frame rotating with the oscillator frequency, keeping only resonant terms in \bref{eq:rabi_coupling}, and  employ the master equation for the density matrix $\hat{\rho}$ of harmonic oscillator plus a single atom ($\hbar=1$)
\begin{align}
\dot{\hat{\rho}}= -\mathrm{i} [\hat{H}(t),\hat{\rho}] + \sum_\alpha {\cal L}_{\hat{L}_\alpha}[\hat{\rho}].
\label{mastereqn}
\end{align}
The Hamiltonian is time dependent due to the classical (uniform) atomic motion \cite{sup:inf}. We include several Lindblad terms  $ {\cal L}_{\hat{O}}[\hat{\rho}]=\hat{O}\hat{\rho}\hat{O}^\dagger - (\hat{O}^\dagger\hat{O}\hat{\rho}+\hat{\rho}\hat{O}^\dagger\hat{O})/2$ accounting for decoherence processes, which are fully described in \cite{sup:inf}:
 (i) Mechanical oscillator states decohere because they are coupled to a heat bath at temperature $T_{\text{osc}}$ with mechanical energy damping rate $\Gamma_{\text{osc}}/(2\pi) = 50 \, \mathrm{Hz}$. Atomic Rydberg states (ii) undergo incoherent relaxation between $\ket{a}$ and $\ket{b}$ due to black body radiation  and (iii)  dephase with a rate $\Gamma_{\text{deph}}/(2\pi) = 1.5 \, \mathrm{kHz}$ due to stray electric fields from the oscillator-bearing surface. The latter effect, a major challenge for Rydberg atom quantum technologies near solid state surfaces, has been steadily reduced \cite{Carter_2013_manipulation,Avigliano_2014_coherence}. (iv) Finally, we incorporate the width of the interrogating atomic beam as a random distribution of initial positions and velocities of a train of Rydberg probe atoms that are all explicitly modelled using \eref{mastereqn}.
 
In our simulations, we consider states $\ket{\astate} = \ket{\nu S_{1/2}, m_J = 1/2}$ and $\ket{\bstate} = \ket{\nu P_{1/2}, m_J = 1/2}$ of $^{87}$Rb with principal quantum number $\nu = 80$. Their resonance frequency is $\omega_{\bstate\astate}/(2\pi) \simeq 6835.81 \, \mathrm{MHz}$ with transition dipole moment $\Vert \mathbf{d}_{\bstate\astate}\Vert \simeq 6711 \,e a_0$ (where $e$ is the electron charge and $a_0$ the Bohr radius). A $148.54 \, \mathrm{nm}$ long and $75.79 \, \mathrm{nm}$ wide CNT with a spherical ferroelectric load can yield a moment of inertia $I \simeq 1.12 \times 10^{-32} \, {\mathrm{kg}}^2 \, \mathrm{m}$ with torsional oscillation frequency $\omega_{\text{osc}}/(2\pi) \simeq 6848.69 \, \mathrm{MHz}$ \cite{sup:inf}, and thus a small atom-oscillator detuning $\delta/(2 \pi) \simeq 12.88 \, \mathrm{MHz}$. A dipole of strength $\Vert \mathbf{d}_{\text{osc}}\Vert \simeq 3.04 \times 10^{9} \, e a_0$ can be attached.
We choose  an impact parameter $D = 21.68 \, \mathrm{\mu m}$, and hence a coupling constant ${\cal K}_0/(2\pi) = 0.64 \, \mathrm{MHz}$.
The transverse atomic wave-guide widths are $\sigma_{X} = \sigma_{Y} = 0.71 \, \mathrm{\mu m}$. The standard deviation of the longitudinal (on-axis) atomic velocity is $\sigma_{v_Z} = 0.1 \, \mathrm{m} \, \mathrm{s}^{-1}$.

To measure phonon quantum numbers in the range $0$-$5$, the corresponding probabilities $P_b$ for the atom to end up in state $\ket{\bstate}$ should be distinguishable, as in the example of \frefp{fig:phaseshifts_displacement}{a}. The beam impact parameter $D$ and atom velocities can be adjusted to yield phase-shifts $\Phi^{(n)}[\delta]$ that fulfil this requirement. The figure demonstrates that even taking into account deviations in $\Phi^{(n)}[\delta]$ due to imperfections as discussed above, a clear inference of $\ket{n}$ can be made. For an oscillator in a Fock state a sequence of QND measurements using atoms can yield the probability $P_b$. If the initial oscillator state $\ket{\Psi}$ is not a Fock state, this sequence initially quickly collapses it into one, say $\ket{n}$, with probability $p_n =|\braket{n}{\Psi}|^2$. A series of such collapse sequences, starting from a re-initialised oscillator state $\ket{\Psi}$ then yields the entire phonon distribution $p_n$ as shown exemplarily in the inset of \frefp{fig:phaseshifts_displacement}{a}.

%
\begin{figure}[htb]
  \begin{centering}
\includegraphics[width= 0.99\columnwidth]{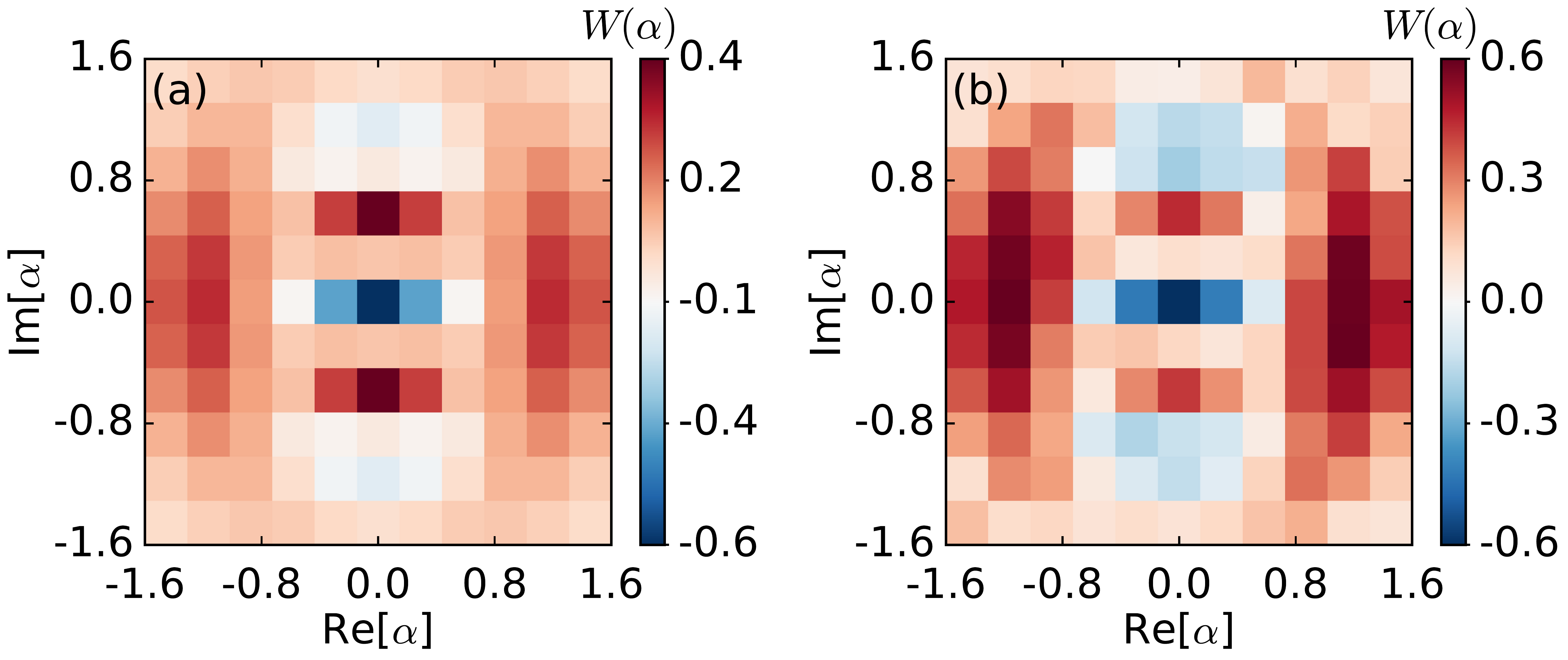} 
\par\end{centering}
\caption{\label{fig:wigner_tomography} 
(Color online) quantum state tomography of oscillator state using QND-detection sequences and coherent displacements all with the same atomic beam. (a) Ideal Wigner-function of the oscillator in state $\ket{\Psi}=[\ket{1} + \ket{3}]/\sqrt{2}$. (b) Reconstruction including practical imperfections as discussed in the text.
}
\end{figure}

\ssection{Quantum state tomography}
%
Phonon-state QND measurements yield the probabilities $p_n = \varrho_{nn} = \mathrm{tr}[\ketbra{n}{n}\hat{\varrho}]$, where $\hat{\varrho}$ is the reduced density matrix for the oscillator, but no coherences between $\ket{n}$, $\ket{m}$.
The full quantum state of the oscillator may be inferred from a tomographical reconstruction of the Wigner function
\begin{equation}
  W(\alpha) = \frac{2}{\pi}\mathrm{tr}[\hat{{\cal D}}^{\dagger}(\alpha)\hat{\varrho} \hat{{\cal D}}(\alpha)\hat{\Pi}],
  \label{eq:wigner_func}
\end{equation}
where  $\hat{{\cal D}}(\alpha) = \exp{[\alpha\hat{c}^{\dagger}-\alpha^{*} \hat{c}]}$ is the displacement operator for a complex amplitude $\alpha$ and $\hat{\Pi} = \mathrm{e}^{\, \mathrm{i}\pi\hat{c}^{\dagger}\hat{c}}$ is the phonon number parity operator.
\eref{eq:wigner_func} is the expectation value of $\hat{\Pi}$ in the state $\hat{\varrho}(-\alpha) = \hat{{\cal D}}(-\alpha)\hat{\varrho} \hat{{\cal D}}^{\dagger}(-\alpha)$. We can thus obtain $W(\alpha)$ as $W(\alpha) = [2/\pi]\sum_n(-1)^n \tilde{p}_n$ from a phonon-distribution $\tilde{p}_n$ as in \frefp{fig:phaseshifts_displacement}{a}, after a coherent displacement by $-\alpha$.

An established method for the quantum coherent displacement of nano-mechanical oscillators does not yet exist. A major advantage of the on-chip architecture proposed here, is that this coherent displacement can be conveniently achieved with the same Rydberg atomic wave guide used for phonon-state measurement.
To this end the atomic dipole transition has to be strongly driven in region $\text{C}$.
Under appropriate conditions, see \cite{sup:inf}, this leads to the effective emergence of a coherent drive for the oscillator.
The evolution operator describing the reduced dynamics of the oscillator for a succession of $N$ atoms reads $\hat{{\cal U}}_{N}(\tau) = \hat{{\cal D}}(\alpha_{N}(\tau))\exp{[\, -\mathrm{i}N \theta(\tau)\hat{c}^{\dagger}\hat{c}]}$, a product of a displacement with complex amplitude $\alpha_{N}(\tau)$ depending on $\mathnormal{\Omega}$ and $\delta$, as well as a phase-shift with $\theta(\tau) = \int^{\tau}{\mathrm{d}t \, {\cal K}^2(\mathbf{R}(t))/\delta}$ that can be compensated (see \cite{sup:inf} for details).
\frefp{fig:phaseshifts_displacement}{b} shows an exemplary oscillator Wigner function before- and after a sequence of $N = 8$ displacement atoms, modelled explicitly as in the previous section. To sample the entire Wigner function with displacements of this kind, one can vary the amplitude and complex phase of the effective Rabi-frequency through parameters of the external drive in region $\text{C}$ \cite{sup:inf}.

To assess the impact of the decoherence sources and imperfections mentioned earlier, we now simulate the complete Wigner tomography sequence:
\begin{enumerate}[(i)]
\item \label {step1} Initialise the oscillator in the state $\hat{\varrho}$ to be measured. This initialisation must be reproducible.
\item \label {step2} Effectuate a coherent displacement, $\hat{\varrho} \mapsto \hat{\varrho}(-\alpha)$, using a flyby sequence of $N$ explicitly modelled displacement atoms.
\item \label {step3} Measure the phonon number with a flyby sequence of $K$ atomic Ramsey interference measurements. The first few atoms collapse the oscillator into a Fock state $\ket{n'}$, which is read out by the remaining majority of the $K$ atoms. 
\item \label {step4} Repeat steps (\ref{step1})-(\ref{step3}) $N_{\text{s}}$ times, to obtain the phonon probability distribution $p_n$ for the displacement $-\alpha$.
\item \label {step5} Repeat step (\ref{step4}) for an ($S \times S$) array of different values for $\alpha\in\mathbb{C}$ to obtain the Wigner function $W(\alpha)$.
 \end{enumerate}
Details on how we implement measurements in our simulation can be found in \cite{sup:inf}.
The Wigner functions reconstructed with this sequence and the ideal expectation are shown in \fref{fig:wigner_tomography} for the  superposition of Fock states $\ket{\Psi}=[\ket{1} + \ket{3}]/\sqrt{2}$. It can be seen that all major qualitative features of the Wigner function, particularly the non-classical negativity, are correctly inferred. Quantitative deviations indicate that the decoherence rates employed here should not be exceeded.

\ssection{Conclusions}
%
By porting technologies from cavity quantum electrodynamics to nano-mechanics, 
our scheme addresses two outstanding challenges for quantum nano-mechanics, namely phonon QND-detection and quantum-state tomography. Thereby, we have also described a technique for the quantum coherent state displacement of nano-mechanical elements. 
The ingredients of our hybrid setup, nano fabricated oscillators and Rydberg atomic waveguides, can naturally co-exist on the same chip surface \cite{keil:atomchip:review}.

\acknowledgments
We gladly acknowledge fruitful discussions with Klemens Hammerer, Swati Singh and Alexander Eisfeld, and EU financial support from the Marie Curie Initial Training Network (ITN) COHERENCE (EU-Grant 265031).
%

\pagebreak
\onecolumngrid
\begin{center}
\textbf{\large Supplemental Material: On-chip quantum tomography of torsional nano-oscillators with guided Rydberg probes}
\end{center}
This supplemental material provides details regarding the complete Hamiltonian, mechanical oscillator displacement via atom driving, and the design criteria for the Rydberg QND probe beam.
\setcounter{equation}{0}
\setcounter{figure}{0}
\setcounter{table}{0}
\makeatletter
\renewcommand{\thesection}{S\arabic{section}}
\renewcommand{\theequation}{S\arabic{equation}}
\renewcommand{\thefigure}{S\arabic{figure}}
\renewcommand{\thetable}{S\arabic{table}}  
\renewcommand{\bibnumfmt}[1]{[S#1]}
\renewcommand{\citenumfont}[1]{S#1}
\makeatother
\section{System Hamiltonian}
\label{sec-hamilt}
\subsection{Torsional oscillator frequency}
The resonance frequency for the harmonic motion of the torsional mechanical oscillator is assumed to be \cite{MePaRo05_1539}
\begin{align}
\omega_{\text{osc}} = \sqrt{\frac{\kappa}{I}} \text{.}
\label{eq:osc_freq}
\end{align}
Here, $I$ is the total moment of inertia with respect to the symmetry axis. It takes into account the entire assembly of ferroelectric particle, and carbon nanotube that comprises the oscillator. The quantity $\kappa$ denotes the torsional spring constant of the nanotube. We consider a value of $\kappa = \SI{2.085e-11}{\newton\meter}$. For a carbon nanotube of mass $m_{\text{cnt}} = \SI{8.71e-19}{\kilo\gram}$ (length $\ell = \SI{148.54}{\nano\meter}$) and diameter $w = \SI{75.79}{\nano\meter}$ with a spherical ferroelectric load of mass $m_{\text{sfl}} = \SI{6.31e-18}{\kilo\gram}$ and radius $r = \SI{63.3}{\nano\meter}$ we obtain a total moment of inertia $I \approx m_{\text{cnt}} w^{2}/4 +  2 m_{\text{sfl}}r^{2}/5 \simeq \SI{1.126e-32}{\kilogram\squared\meter}$. This finally corresponds to a frequency $\omega_{\text{osc}}/(2 \pi) \simeq \SI{6848.69}{\mega\hertz}$ as specified in the main text.

\subsection{Atom-oscillator coupling}
\begin{figure}[htb]
  \centering
    \includegraphics[width=0.35\linewidth]{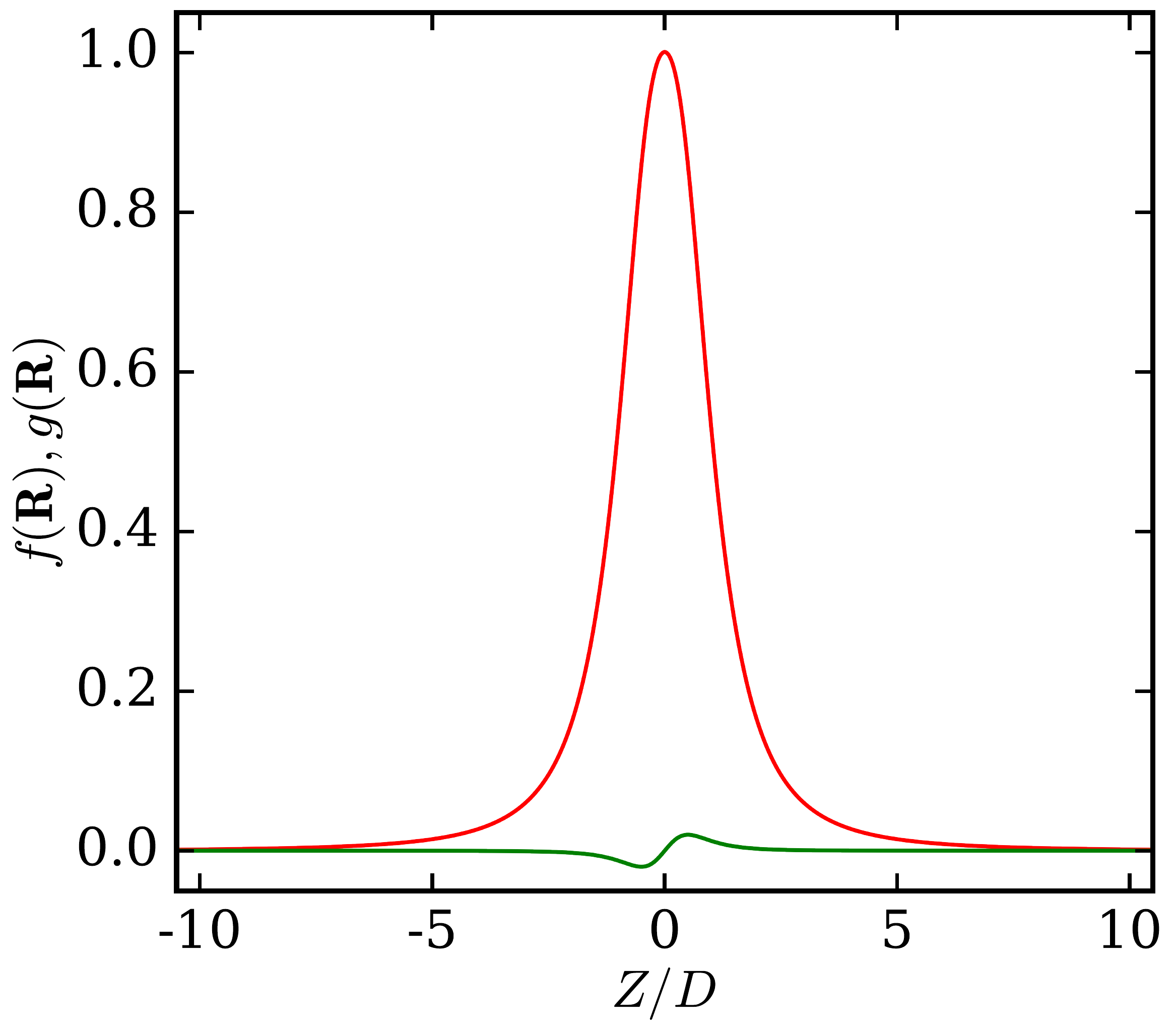}
\caption{\label{fig:couplingfuncs}
Coupling functions $f(X,Y,Z)$ (red) and $g(X,Y,Z)$ (green) along the axis of the waveguide. Parameters are $D \simeq \SI{21.68}{\micro\metre}$, $X/D \simeq 0.0231$, and $Y/D =  0.9997$.
}
\end{figure}

The atoms passing by the oscillator experience a dipole-dipole interaction
\begin{align}
\hat{H}_{\text{int}}&=\frac{1}{4\pi\varepsilon_{0}R^{3}}[\mathbf{d}_{\bstate \astate}\cdot\mathbf{d}_{\text{osc}}-3(\mathbf{d}_{\text{osc}}\cdot\mathbf{u})(\mathbf{d}_{\bstate \astate}\cdot\mathbf{u})]\left[\hat{\sigma}_{\bstate \astate}+\hat{\sigma}_{\astate \bstate}\right]\text{,}
\label{eq:dip_dipint}
\end{align}
with the permanent dipole-moment $\mathbf{d}_{\text{osc}}$ of the nano-particle attached to the oscillator. Here the vector $\mathbf{R}$ points from the oscillator to the atom and $\mathbf{u}=\mathbf{R}/R$ with $R = \Vert\mathbf{R}\Vert$. The direction of the atomic transition dipole-moment $\mathbf{d}_{\bstate \astate}$ in principle depends on the states in question. For simplicity we choose the two atomic states such that their transition dipole-moment is along the $z$-axis \cite{footnote:transitiondipolefixed}. In this case, the interaction reduces to
\begin{align}
\hat{H}_{\text{int}}(\mathbf{R}, \varphi)&=\frac{d_{\bstate \astate} d_{\text{osc}}}{4\pi\varepsilon_{0}R^{3}}\bigg[\Big(1 - 3\frac{Z^2}{R^2}\Big) \cos \varphi - 3\frac{X Z}{R^2} \sin \varphi \bigg]\bigg[\hat{\sigma}_{\bstate \astate} + \hat{\sigma}_{\astate \bstate}\bigg]
  \CR
  &=V_0\Big[f(\mathbf{R}) \cos \varphi - g(\mathbf{R}) \sin \varphi \Big]\Big[\hat{\sigma}_{\bstate \astate} + \hat{\sigma}_{\astate \bstate}\Big]\text{,}
  \label{eq:full_interaction}
\end{align}
with $d_{\text{osc}}=\Vert \mathbf{d}_{\text{osc}}\Vert $, $d_{\bstate \astate}=\Vert \mathbf{d}_{\bstate \astate}\Vert $, $V_0 = \frac{d_{\bstate \astate} d_{\text{osc}}}{4\pi\varepsilon_{0}D^{3}}$, $\mathbf{R} = (X,Y,Z)$  and coupling functions  $f(\mathbf{R}) = [D/R]^3[1 - 3Z^2/R^2]$, $g(\mathbf{R}) = 3 X Z D^3/R^5$. These are sketched for an example in \fref{fig:couplingfuncs}.
We now make use of the fact that the coordinate $\varphi$ is harmonically bound to a stable equilibrium configuration given by the angle $\varphi_0 = \pi/2$, and introduce the small angular displacement $|\delta \varphi| \ll \varphi_0$, such that $\varphi = \varphi_0 + \delta \varphi$.

Then a Taylor series expansion of the sinusoidal functions in the potential energy, around the equilibrium value $\varphi_0 = \pi/2$, and up to second order in $\delta \varphi$ yields
\begin{align}
\hat{H}_{\text{int}}(\mathbf{R}, \varphi)&\approx V_0\Big[g(\mathbf{R}){\delta \varphi}^2 - f(\mathbf{R})\delta \varphi - g(\mathbf{R}) \Big]\Big[\hat{\sigma}_{\bstate \astate} + \hat{\sigma}_{\astate \bstate}\Big]\text{.}
\end{align}
Next, we recast the above in terms of the mechanical phonon creation and annihilation operators $\hat{c}$ and $\hat{c}^{\dagger}$ as
\begin{align}
  &{\delta \varphi} = \varphi_{\text{zpm}}[\hat{c} + \hat{c}^{\dagger}]
  \CR
  &{\delta \varphi}^2 = \varphi_{\text{zpm}}^2[\hat{c}^2 + (\hat{c}^{\dagger})^2 + 2\hat{c}^{\dagger}\hat{c} + \hat{\openone}_{\text{osc}}]\text{.}
\end{align}
where $\varphi_{\text{zpm}} = \sqrt{\hbar/(2 \omega_{\text{osc}} I)}$ represents the amplitude of the zero point motion of the torsion oscillator and $\hat{\openone}_{\text{osc}} = \sum_n\ketbra{n}{n}$.

For atoms travelling perfectly at the centre of the atomic waveguide ($X=0$, see Fig.~1 in the main article), we then obtain the interaction Hamiltonian $\hat{H}_{\text{int}}$ given in the main article.

\section{Ramsey measurements of phase shifts}
\label{Ramsey}
As discussed in the main text, the train of Rydberg atoms passing the oscillator is modelled explicitly, with atom $k$ being given a randomized initial position $\bv{R}_k(0)$ and velocity $\bv{v}_k$, subsequently following a uniform trajectory $\bv{R}_k(t) = \bv{R}_k(0) + \bv{v}_k t$. The initial widths of these random distributions, $\sigma_{X,Y}$ in the position plane transverse to the beam and $\sigma_{v_{Z}}$ in the velocity along the beam, are chosen to mimic the relevant uncertainties for a beam of atoms travelling within a very tight waveguide. Uniform motion is justified as long as forces on the atom are negligible or weak compared to waveguide trapping, as we assume here.

To implement a single interferometric measurement we excert a sequence of two identical microwave pulses (mw) onto an atom.
The two mw pulses are applied in $\text{R}_1$ and $\text{R}_2$, and thus they are delayed from each other by a time period $\tau = L/|\mathbf{v}|$, which is the time of flight of the atom in the region $\text{C}$ of length $L = \SI{30.872}{\micro\metre}$, during which the atom is let to interact with the mechanical oscillator.
We model the atomic evolution in the two locations $\text{R}_{1/2}$ by applying a unitary transformation $\hat{A}_{\pi/2}(\phi)$, that in the basis $\{\ket{\astate}, \ket{\bstate}\}$ reads
\begin{align}
  & A_{\pi/2}(\phi) = \frac{1}{\sqrt{2}}
    \begin{pmatrix}
    1 & -\mathrm{e}^{\, \mathrm{i}\phi} \\
    \mathrm{e}^{\, -\mathrm{i}\phi} & 1\\
  \end{pmatrix} \text{.}
\label{eq:rabi_pulse}
\end{align}
Here $\phi$ represents a phase that in practice can be controlled by tuning the microwave pulse. In an experiment one adjusts the relative phase between the two pulses to scan the Ramsey fringes of the signals described by the probabilities $P_{\bstate}$ and $P_{\astate}$ that the atom is detected in the state $\ket{\bstate}$ or $\ket{\astate}$, respectively, when reaching the detector at F.
We use $\hat{A}_{\pi/2}(\phi=0)$ in region $\text{R}_{1}$ and choose the phase of $\hat{A}_{\pi/2}(\phi)$ in region $\text{R}_2$ such that $P_{\bstate}$ equals zero when the mechanical oscillator mode is in its ground state and the atom has traversed the interferometer with velocity and transverse position corresponding to the mean atomic beam values. 

Finally, the relative phase shift $\Phi^{({n})}[\delta]$ incurred by each atom depends on the difference between two adjacent eigenenergies $E_{-}^{(n+1)}(\mathbf{R}) - E_{+}^{(n)}(\mathbf{R})$ integrated over time and thus can be found from $E_{\pm}^{(n)}(\mathbf{R}) = \hbar\mathnormal{\delta}/2 \Big[-1 \pm \sqrt{1 + 4 n{\cal K}^2(\mathbf{R})/\mathnormal{\delta}^2}\Big]$ following \cite{Brune_1990_nondemolition}.
This helps to choose the right parameters for mapping the relative phase shift $\Phi^{({n})}[\delta]$ from a selected range of phonon numbers (e.g.~$0$-$5$) onto the interval $[0,\pi]$ (see \sref{param_choices}).

\section{Effective coherent driving of mechanical oscillator}
\label{displacement}
Simultaneous action of the Hamiltonians $\hat{H}_{\text{int}}$ and $\hat{H}_{\text{coup}}$ from the main article can effectively create a drive for the quantum harmonic oscillator. For that we need to have $|\mathnormal{\delta}| > |\mathnormal{\Omega}(t)|, |{\cal K}(\mathbf{R})|$ and start with all the atomic population in $\ket{a}$. In this regime we can adiabatically eliminate the second Rydberg state $\ket{\bstate}$ to obtain the effective Hamiltonian $\hat{H}_{\text{dho}}(t)\otimes\hat{\sigma}_{\astate \astate}$, where
\begin{align}
  & \hat{H}_{\text{dho}}(t) = \frac{{\cal K}^{2}(\mathbf{R})}{\mathnormal{\delta}}\hat{c}^{\dagger}\hat{c} - \frac{{\cal K}(\mathbf{R})\mathnormal{\Omega}(t)}{2 \mathnormal{\delta}}\hat{c} - \frac{{\cal K}(\mathbf{R})\mathnormal{\Omega}^{*}(t)}{2 \mathnormal{\delta}}\hat{c}^{\dagger}.
\label{eq:Hfho}
\end{align}
While the atom essentially remains in the Rydberg state $\ket{\astate}$, the evolution for the mechanical oscillator can be written as 
\begin{align}
  &\hat{\varrho}(\tau) =\hat{{\cal U}}(\tau) \hat{\varrho}(0) \hat{{\cal U}}^{\dagger}(\tau),
\label{eq:oscevolution}
\end{align}
where $ \hat{\varrho}$ denotes the density matrix describing the oscillator. The quantity $\hat{{\cal U}}(\tau)$ equals the time development operator for a driven quantum harmonic oscillator, which one can see by exploiting the commutation relations of $\hat{c}$, $\hat{c}^{\dagger}$ \cite{Gardiner_2004_noise}
\begin{align}
  &\hat{{\cal U}}(\tau) =\mathrm{e}^{\, \mathrm{i} \lambda (\tau)} \, \hat{{\cal D}}\Big(\xi(\tau)\mathrm{e}^{\, -\mathrm{i}\theta(\tau)}\Big) \, \mathrm{e}^{\, -\mathrm{i} \theta(\tau)\hat{c}^{\dagger}\hat{c}},
  \CR
  & \lambda(\tau) = -\frac{1}{2 \hbar^2}\int_{0}^{\tau}{\mathrm{d}t\int_{0}^{t}{\mathrm{d}t' \big[\hat{H}_{\text{dho}}(t), \hat{H}_{\text{dho}}(t')\big]}},
  \CR
  & \xi(\tau) =  \mathrm{i} \int^{\tau}{\mathrm{d}t \frac{\mathnormal{\Omega}^{*}(t){\cal K}(t)}{2 \mathnormal{\delta}}\mathrm{e}^{\, \mathrm{i}\theta(t)}},
  \CR
  & \theta(\tau) = \int^{\tau}{\mathrm{d}t \frac{{\cal K}^2(t)}{\mathnormal{\delta}}}.
\label{eq:Uosc}
\end{align}
Here we adopted the shortened notation ${\cal K}(t) = {\cal K}(\mathbf{R}(t))$ and $\hat{{\cal D}}$ is the displacement operator introduced in the main text. Since $\lambda(\tau)$ is a $c$-number, $\exp{[\mathrm{i} \lambda(\tau)]}$ is a global phasor that we will ignore from now on.
The $N$-th power of $\hat{{\cal U}}(\tau)$ then accounts for the state evolution of the mechanical oscillator after a successive fly-by of $N$ atoms, each atom passing through the oscillator in a time interval $\tau$.
To compute $\hat{{\cal U}}_N(\tau) \equiv [\hat{{\cal U}}(\tau)]^{N}$ we use the following properties of the displacement operator:
\begin{align}
  & \hat{{\cal D}}(\alpha)\hat{{\cal D}}(\beta) = \exp{[(\alpha \beta^{*} - \alpha^{*}\beta)/2]}\hat{{\cal D}}(\alpha + \beta),
  \CR
  & \exp{[\mathrm{i} \theta \hat{c}^{\dagger}\hat{c}]}\hat{{\cal D}}(\alpha) = \hat{{\cal D}}(\alpha \exp{[\mathrm{i} \theta]})\exp{[\mathrm{i} \theta \hat{c}^{\dagger}\hat{c}]} \text{,}
\label{eq:Dproperties}
\end{align}
such that, ignoring again global phasors, one has
\begin{align}
  & \hat{{\cal U}}_N(\tau) = \hat{{\cal D}}\Big(\xi(\tau)\sum_{l=1}^{N}\mathrm{e}^{\, -\mathrm{i}l\theta(\tau)}\Big) \, \mathrm{e}^{\, -\mathrm{i}N \theta(\tau)\hat{c}^{\dagger}\hat{c}}= \hat{{\cal D}}\big(\alpha_N(\tau)\big) \, \mathrm{e}^{\, -\mathrm{i}N \theta(\tau)\hat{c}^{\dagger}\hat{c}},
  \CR
  & \alpha_N(\tau) = \frac{\sin \big(N \theta(\tau)/2\big)}{\sin \big(\theta(\tau)/2\big)}\xi(\tau)\mathrm{e}^{\, -\mathrm{i}[N + 1]\theta(\tau)/2}.
\label{eq:UoscMatoms}
\end{align}
In the manuscript, all the numerical calculations involving the Hamiltonian $\hat{H}_{\text{coup}}$ were assuming continuous waves with $\mathnormal{\Omega}(t) = \mathnormal{\Omega}_0$.
Sampling the dynamical phase space of the mechanical oscillator is then achieved by adjusting the amplitude and phase of the complex Rabi frequency $\mathnormal{\Omega}_0$, while taking into account the additional phase offset generated by $\exp{[\, -\mathrm{i}N \theta(\tau)\hat{c}^{\dagger}\hat{c}]}$.

\section{Decoherence sources}
\label{lindblad}

As discussed in the main text, we consider a variety of practically relevant decoherence sources, in order to explore the limitation of our proposal.

The specific Lindblad operators with which we describe the effects listed in the main text are as follows: (i) Mechanical oscillator states decohere because they are coupled to a heat bath equilibrated at a temperature $T_{\text{osc}}$. This is described by the two terms $\hat{L}_{-}=\sqrt{(\bar{n}_{\text{th}} + 1)\Gamma_{\text{osc}}}\hat{c}$ and $\hat{L}_{+}=\sqrt{\bar{n}_{\text{th}}\Gamma_{\text{osc}}}\hat{c}^\dagger$, with a thermal occupation number $\bar{n}_{\text{th}} = (\exp{[\hbar \omega_{\text{osc}}/(k_{\text{B}} T_{\text{osc}})]} - 1)^{-1}$ and a mechanical energy damping rate $\Gamma_{\text{osc}} = \omega_{\text{osc}}/Q_{\text{osc}}$ for a given quality factor $Q_{\text{osc}}$ of the mechanical oscillator. (ii) Atomic Rydberg states are assumed to undergo pure relaxation due to black body radiation induced transitions, modelled with two terms $\hat{L}_{\mu'\mu}=\sqrt{\Gamma_{\text{bbr}}}\hat{\sigma}_{\mu'\mu}$, where $\{\mu',\mu\}=\{a,b\}$ or $\{b,a\}$. We employ $\Gamma_{\text{bbr}}/(2\pi)=\SI{988.63}{\hertz}$, determined following \cite{Beterov_2009_calculations}. (iii) They are also assumed subject to dephasing with $\hat{L}_{\text{deph},\mu}=\sqrt{\Gamma_{\text{deph}}}\hat{\sigma}_{\mu \mu}$ using $\mu\in\{a,b \}$ due to stray electric fields from the oscillator-bearing surface \cite{Carter_2013_manipulation}. We employ $\Gamma_{\text{deph}}/(2\pi)= \SI{1.50}{\kilo\hertz}$, the same order of magnitude as values reported in \cite{Avigliano_2014_coherence}.

Due to the short distances between Rydberg probes and mechanical oscillator, a primary challenge for our scheme are the finite widths of the atomic beam within the waveguide. The implementation of the interrogating atomic beam in the waveguide via a random distribution of initial positions $\bv{R}_k(0)$ and velocities $\bv{v}_k$ was discussed in \sref{Ramsey}. 

\section{Quantum state tomography simulations}

\ssection{Simulations of measurement sequences}
\label{tomography_simulation}
%
The physical sequence for a full oscillator quantum state tomography has been discussed in the main text. Here we supply further technical details about the simulations.
In the following let $\hat{\rho}$ denote the full density matrix of the compound atom-oscillator system. We then use $\hat{\varrho} = \mathrm{tr}_{\text{at}}[\hat{\rho}]$
for the reduced density operator of the oscillator after tracing over the atomic degrees of freedom.

Each atom, whether for a Ramsey measurement or a phase space displacement of the oscillator state, is initialized in $\ket{a}$ and made to move on a trajectory as discussed in \sref{Ramsey}. Then its fly-by past the oscillator is modelled with the master equation (3) of the main text.
The mw pulses required for Ramsey measurements are emulated via the instantaneous application of the operator $\hat{A}_{\pi/2}(\phi)$ in \eref{eq:rabi_pulse}.
The Lindblad terms listed in \sref{lindblad} take into account decoherence processes. Of crucial importance for modelling the experimental sequence is the final atom state detection at $\text{F}$. We assume detection can only yield the two states $\ket{\astate}$ or $\ket{\bstate}$.
To numerically represent this measurement, we compare a pseudorandom number $\eta$, drawn from a standard uniform distribution, with the probability $P_{\bstate} = \mathrm{tr}[\hat{\sigma}_{\bstate \bstate} \hat{\rho}]$ that the atom is found in $\bstate$.
The output of the measurement is $\hat{\sigma}_{\astate \astate}\hat{\rho}\hat{\sigma}_{\astate \astate}/[1 - P_{\bstate}]$ if $P_{\bstate} < \eta$ and $\hat{\sigma}_{\bstate \bstate}\hat{\rho}\hat{\sigma}_{\bstate \bstate}/P_{\bstate}$ otherwise, thus collapsing the state onto $\ket{\astate}$ or $\ket{\bstate}$ in the subspace of the atom \cite{Wiseman_2009_measurement}.
After a series of $K = 43$ atoms typically all but one of the phonon probabilities are depopulated, such that $\varrho_{nn} \simeq \mathrm{\delta}_{n m}$, with $\mathrm{\delta}_{n m}$ being the kronecker delta and $m$ a positive integer, the final phonon number.
The mechanical oscillator is then assumed to have collapsed into the a priori unknown Fock state $\ket{m}$. In the theory, we can directly extract $m$ from the simulation, repeat the process multiple times and thus extract the entire phonon distribution. We call this approach ``Method A''.

However, an experiment would not have access to $m$ directly, instead it would extract the probabilities $P_{a/b}$ from the measurement results of the $K = 43$ probing atoms. Using Fig.~2 (a) of the main article, these can then be translated into values of $m$, but this translation may be subject to different error sources. We also extract a second value of $m$ from the simulation in this manner, called ``Method B''. 

We finally sample the entire phase space of the oscillator on a square grid of $S \times S$ points, by explicitly modelling different displacements $\alpha\in\mathbb{C}$.
The density matrix is propagated in time using Eq.~(3) of the main text, while the passing atoms are driven, such that it effectively evolves as described by the operator $\hat{{\cal U}}_{N}$.
This is described in \sref{displacement} and visualized in Fig.~2 (b) of the main article. For each point of the square grid a phonon count distribution is sampled and the Wigner function is finally computed as discussed in the main text.

\begin{figure}[htb]
  \centering
   \includegraphics[width=0.6\linewidth]{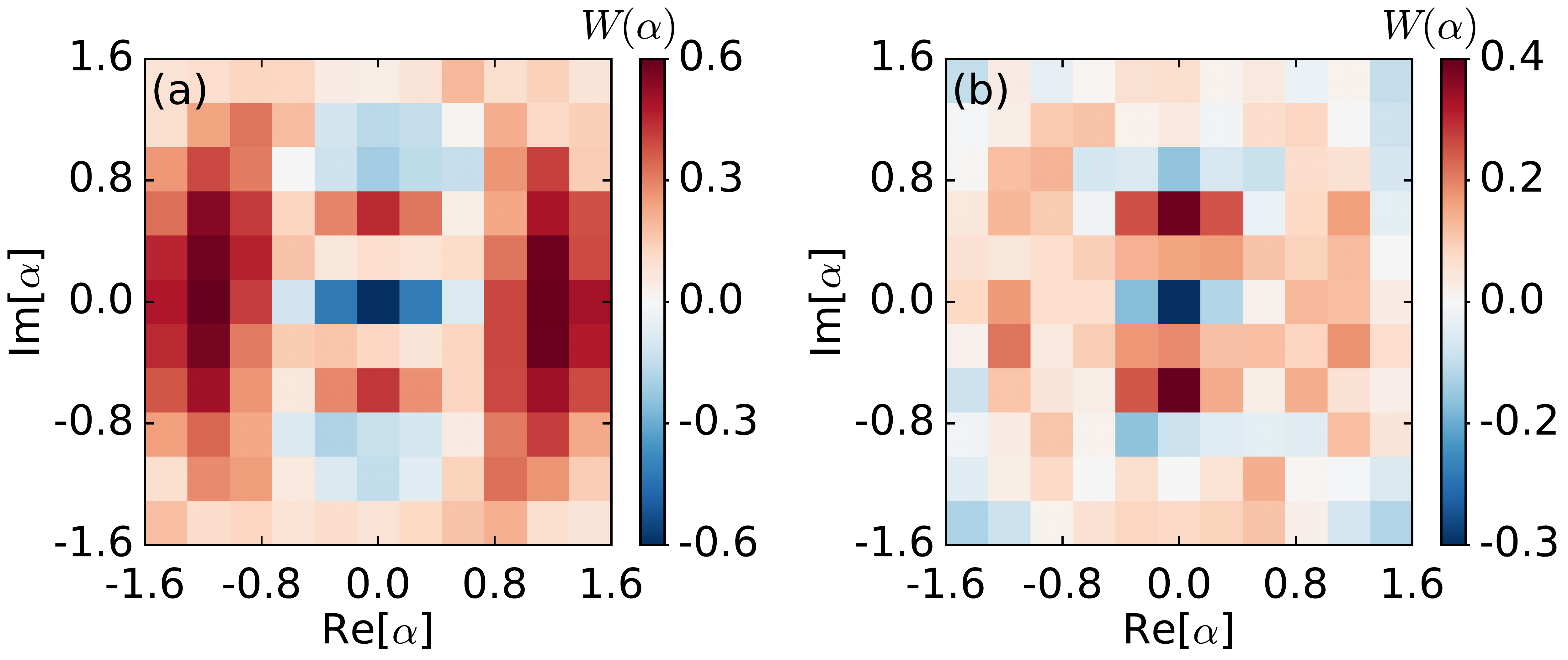}
\caption{\label{fig:tomography_methods}
Tomographical reconstruction of the Wigner density $W(\alpha)$ of the mechanical oscillator superposition state $\ket{\Psi}=[\ket{1} + \ket{3}]/\sqrt{2}$. (a) Evaluation of $W(\alpha)$ using the reconstruction protocol of ``Method A''. (b) The outcome of the same $W(\alpha)$ applying instead the protocol of ``Method B''. Both, ``Method A'' and ``Method B'', are defined in \sref{tomography_simulation}.
}
\end{figure}

Figure~3 (b) of the main text illustrates the Wigner density of the mechanical oscillator superposition state $\ket{\Psi}=[\ket{1} + \ket{3}]/\sqrt{2}$ as obtained through ``Method A''.
Here we show it again in \frefp{fig:tomography_methods}{a} together with \frefp{fig:tomography_methods}{b}, which depicts also the Wigner density of $\ket{\Psi}=[\ket{1} + \ket{3}]/\sqrt{2}$ derived in this case from ``Method B''.
The Wigner density computed via ``Method A'' resembles the exact outcome displayed in Fig.~3 (a) of the main text more than the counterpart result determined through ``Method B''.
Considering that occasionally a series of $K = 43$ Ramsey measurements may not suffice to project the mechanical oscillator state into a Fock state, we expect a higher inaccuracy of ``Method B'' compared to ``Method A''.
Indeed, if a complete Fock state collapse is not realized, applying ``Method B'' may lead to a wrong Fock state record or to the loss of a statistical sample.
Contrarily, if we apply ``Method B'' in a similar situation, a statistical loss never occurs and the error for a Fock state miscount is lower.
%
\section{System design and parameter choices}
\label{param_choices}
The following central requirements dictate the choice of parameters for the setup:
\begin{itemize}
\item The distance $L$ between regions $\text{R}_{1,2}$ and the parameters $D$ and $\bv{v}$ have to be adjusted such that
(i) the average fly by time $\tau$ of an atom across the interferometer is much shorter than the Rydberg lifetime, and (ii) each atomic record can serve as a non-destructive measurement of discrete (phonon) number states in the range between $n = 0$ and $n = 5$, as can be seen from the atomic excited state probability $P_{\bstate}$ shown in Fig.~2 (a) of the main article.
\item For each detection event there is no energy exchange between the atom and the mechanical oscillator.
In other words, the instantaneous atom-oscillator coupling is designed to remain sufficiently weak (off-resonant) until the atom interferometric measurement is completed.
In this way, transitions between atomic states $\ket{\bstate}$ and $\ket{\astate}$ via the absorption or emission of a phonon are negligible, such that the mechanical oscillator contains $n$ phonons before and after the atom traverses the interferometer.
\end{itemize}
%
\begin{table}
  {\def\arraystretch{1.6}\tabcolsep=3.6pt
  \begin{center}
  \begin{tabular}{|p{7.0cm}|c|c|c}
  \hline
  \textbf{Atomic system (Rubidium, $^{87}$Rb)} & Symbol & Value \\ \hline \hline 
  Mass & $M$ & $\SI{1.44e-25}{\kilogram}$  \\
  Initial atom position (with $D$ being the impact parameter)& $\mathbf{R}(t=0) = (X(0),Y(0)=D,Z(0))$ & $(\SI{0.0}{\micro\metre},\SI{21.675}{\micro\metre},\SI{-15.436}{\micro\metre})$ \\
  Spatial atomic beam widths (transverse to $z$-axis) & $\sigma_X = \sigma_Y$ & $\SI{0.707}{\micro\metre}$ \\
  \begin{tabular}[t]{@{}c@{}}Initial atom velocity: displacement sequence \\ \phantom{Initial atom velocity: }measurement sequence\end{tabular} & $\mathbf{v}(t=0) = (v_X(0), v_Y(0), v_Z(0))$ & \begin{tabular}[t]{@{}c@{}} $(\SI{0}{\metre\per\second},\SI{0}{\metre\per\second},\SI{14}{\metre\per\second})$ \\ $(\SI{0}{\metre\per\second},\SI{0}{\metre\per\second},\SI{8}{\metre\per\second})$\end{tabular} \\
  Velocity atomic beam width (along $z$-axis) & $\sigma_{v_Z}$ & \SI{0.01}{\meter\per\second} \\
  Principal quantum number & $\nu$ & 80 \\
  Rydberg state basis & $\{\ket{\nu L_J, m_J}\}$ & $\{\ket{\astate} = \ket{80 S_{1/2}, 1/2}, \ket{\bstate} = \ket{80 P_{1/2}, 1/2}\}$ \\
  Transition frequency ($\ket{b}\leftrightarrow\ket{a}$) & $\omega_{\bstate \astate}/(2\pi)$ & $\SI{6835.81}{\mega\hertz}$ \\
  Electric dipole moment strength & $d_{\bstate \astate}$ & $\SI{5.69e-26}{\coulomb\metre}$ \\ \hline \hline 
  \textbf{Torsional mechanical oscillating mode} & & \\ \hline \hline 
  Torsional spring constant of the nanotube & $\kappa$ & $\SI{2.085e-11}{\newton\meter}$\\
  Total moment of inertia with respect to the tube axis & $I$ & $\SI{1.126e-32}{\kilogram\squared\meter}$ \\
  Permanent dipole moment strength of ferroelectric load & $d_{\text{osc}}$ & $\SI{2.58e-20}{\coulomb\metre}$ \\
  Frequency & $\omega_{\text{osc}}/(2\pi) = (2\pi)^{-1}\sqrt{\kappa/I}$ & $\SI{6848.69}{\mega\hertz}$ \\
  Number state basis & $\{\ket{n}\}$ & $\{\ket{0}, \ket{1}, \dots, \ket{15}\}$ \\
  Quality factor & $Q = \omega_{\text{osc}}/\Gamma_{\text{osc}}$ & $1.37 \cdot 10^{8}$ \\ 
  Heat bath temperature & $T_{\text{osc}}$ & $\SI{0.025}{\kelvin}$ \\ \hline \hline
  \textbf{Coupling and decoherence rates} & & \\ \hline \hline 
  \rule{0pt}{4ex}Atom-oscillator coupling rate & ${\cal K}_0/(2\pi) = \dfrac{d_{\bstate \astate} d_{\text{osc}}}{8\pi^2\varepsilon_{0}D^{3}}\dfrac{1}{\sqrt{2 \hbar \omega_{\text{osc}} I}}$ & $\SI{0.64}{\mega\hertz}$ \\
  Effective Rabi frequency & ${\Omega_0}/(2\pi)$ & \SIrange{0.0}{1.8}{\mega\hertz} \\
  Mechanical damping rate & $\Gamma_{\text{osc}}/(2\pi)$ & $\SI{50}{\hertz}$ \\
  Pure relaxation rate due to black body radiation induced transitions ($\ket{b}\leftrightarrow\ket{a}$) & $\Gamma_{\text{bbr}}/(2 \pi)$ & $\SI{988.63}{\hertz}$ \\
  Pure dephasing rate of $\ket{\astate}$ and $\ket{\bstate}$ levels due to noisy stray electric fields & $\Gamma_{\text{deph}}/(2\pi)$ & $\SI{1.50}{\kilo\hertz}$ \\ \hline \hline
  \textbf{Protocol of state tomography} & & \\ \hline \hline 
  Dimensions (number of pixels) of the reconstructed Wigner function  & $S \times S$ & $11 \times 11$ \\
  Number of atoms per displacement sequence to reach a given phase space pixel & $N$ & 8 \\
  Number of atoms per measurement sequence to collapse oscillator into Fock state& $K$ & 43 \\
  Number of repetitions (samples) of a displacement plus measurement sequence to obtain a set of phonon probabilities at a given pixel & $N_{\text{s}}$ & $512$ \\
  Atom-oscillator detuning & $|\delta|/(2 \pi) = |\omega_{\text{osc}} - \omega_{\bstate \astate}|/(2 \pi)$ &  $\SI{12.88}{\mega\hertz}$ \\
  Passage time per atom in a displacement sequence & $\tau_{\mathrm{disp}}$ & $\SI{2.205}{\micro\second}$ \\
  Passage time  per atom in a measurement sequence & $\tau_{\mathrm{meas}}$ & $\SI{3.859}{\micro\second}$
 \\ \hline
\end{tabular}%
\end{center}%
}
\caption{Parameters used for our simulations underlying Fig.~2 and Fig.~3 of the main article.
\label{parameters}}
\end{table}


\begin{thebibliography}{45}
\expandafter\ifx\csname natexlab\endcsname\relax\def\natexlab#1{#1}\fi
\expandafter\ifx\csname bibnamefont\endcsname\relax
  \def\bibnamefont#1{#1}\fi
\expandafter\ifx\csname bibfnamefont\endcsname\relax
  \def\bibfnamefont#1{#1}\fi
\expandafter\ifx\csname citenamefont\endcsname\relax
  \def\citenamefont#1{#1}\fi
\expandafter\ifx\csname url\endcsname\relax
  \def\url#1{\texttt{#1}}\fi
\expandafter\ifx\csname urlprefix\endcsname\relax\def\urlprefix{URL }\fi
\providecommand{\bibinfo}[2]{#2}
\providecommand{\eprint}[2][]{\url{#2}}

\bibitem[{\citenamefont{Abbott et~al.}(2016)\citenamefont{Abbott, Abbott,
  Abbott, Abernathy, Acernese, Ackley, Adams, Adams, Addesso, Adhikari
  et~al.}}]{abbot_gravwave_PRL}
\bibinfo{author}{\bibfnamefont{B.~P.} \bibnamefont{Abbott}},
  \bibinfo{author}{\bibfnamefont{R.}~\bibnamefont{Abbott}},
  \bibinfo{author}{\bibfnamefont{T.~D.} \bibnamefont{Abbott}},
  \bibinfo{author}{\bibfnamefont{M.~R.} \bibnamefont{Abernathy}},
  \bibinfo{author}{\bibfnamefont{F.}~\bibnamefont{Acernese}},
  \bibinfo{author}{\bibfnamefont{K.}~\bibnamefont{Ackley}},
  \bibinfo{author}{\bibfnamefont{C.}~\bibnamefont{Adams}},
  \bibinfo{author}{\bibfnamefont{T.}~\bibnamefont{Adams}},
  \bibinfo{author}{\bibfnamefont{P.}~\bibnamefont{Addesso}},
  \bibinfo{author}{\bibfnamefont{R.~X.} \bibnamefont{Adhikari}},
  \bibnamefont{et~al.} (\bibinfo{collaboration}{LIGO Scientific Collaboration
  and Virgo Collaboration}), \bibinfo{journal}{Phys. Rev. Lett.}
  \textbf{\bibinfo{volume}{116}}, \bibinfo{pages}{061102}
  (\bibinfo{year}{2016}).

\bibitem[{\citenamefont{Chan et~al.}(2011)\citenamefont{Chan, Alegre,
  {Safavi-Naeini}, Hill, Krause, Gr{\"o}blacher, Aspelmeyer, and
  Painter}}]{chan:groundstatecooling}
\bibinfo{author}{\bibfnamefont{J.}~\bibnamefont{Chan}},
  \bibinfo{author}{\bibfnamefont{P.~M.} \bibnamefont{Alegre}},
  \bibinfo{author}{\bibfnamefont{A.~H.} \bibnamefont{{Safavi-Naeini}}},
  \bibinfo{author}{\bibfnamefont{J.~T.} \bibnamefont{Hill}},
  \bibinfo{author}{\bibfnamefont{A.}~\bibnamefont{Krause}},
  \bibinfo{author}{\bibfnamefont{S.}~\bibnamefont{Gr{\"o}blacher}},
  \bibinfo{author}{\bibfnamefont{M.}~\bibnamefont{Aspelmeyer}},
  \bibnamefont{and} \bibinfo{author}{\bibfnamefont{O.}~\bibnamefont{Painter}},
  \bibinfo{journal}{Nature} \textbf{\bibinfo{volume}{478}}, \bibinfo{pages}{89}
  (\bibinfo{year}{2011}).

\bibitem[{\citenamefont{Aspelmeyer et~al.}(2014)\citenamefont{Aspelmeyer,
  Kippenberg, and Marquardt}}]{aspelmeyer:review}
\bibinfo{author}{\bibfnamefont{M.}~\bibnamefont{Aspelmeyer}},
  \bibinfo{author}{\bibfnamefont{T.~J.} \bibnamefont{Kippenberg}},
  \bibnamefont{and}
  \bibinfo{author}{\bibfnamefont{F.}~\bibnamefont{Marquardt}},
  \bibinfo{journal}{Rev. Mod. Phys.} \textbf{\bibinfo{volume}{86}},
  \bibinfo{pages}{1391} (\bibinfo{year}{2014}).

\bibitem[{\citenamefont{Kippenberg and Vahala}(2007)}]{kippenberg:review}
\bibinfo{author}{\bibfnamefont{T.}~\bibnamefont{Kippenberg}} \bibnamefont{and}
  \bibinfo{author}{\bibfnamefont{K.}~\bibnamefont{Vahala}},
  \bibinfo{journal}{Opt. Express} \textbf{\bibinfo{volume}{15}},
  \bibinfo{pages}{17172} (\bibinfo{year}{2007}).

\bibitem[{\citenamefont{Poot and {van der Zant}}(2012)}]{poot:mechquantumsyst}
\bibinfo{author}{\bibfnamefont{M.}~\bibnamefont{Poot}} \bibnamefont{and}
  \bibinfo{author}{\bibfnamefont{H.~S.} \bibnamefont{{van der Zant}}},
  \bibinfo{journal}{Phys. Rep.} \textbf{\bibinfo{volume}{511}},
  \bibinfo{pages}{273} (\bibinfo{year}{2012}).

\bibitem[{\citenamefont{Palomaki et~al.}(2013)\citenamefont{Palomaki, Teufel,
  Simmonds, and Lehnert}}]{Palomaki_osc_mw_entanglement_science}
\bibinfo{author}{\bibfnamefont{T.~A.} \bibnamefont{Palomaki}},
  \bibinfo{author}{\bibfnamefont{J.~D.} \bibnamefont{Teufel}},
  \bibinfo{author}{\bibfnamefont{R.~W.} \bibnamefont{Simmonds}},
  \bibnamefont{and} \bibinfo{author}{\bibfnamefont{K.~W.}
  \bibnamefont{Lehnert}}, \bibinfo{journal}{Science}
  \textbf{\bibinfo{volume}{342}}, \bibinfo{pages}{710} (\bibinfo{year}{2013}).

\bibitem[{\citenamefont{Riedinger et~al.}(2016)\citenamefont{Riedinger, Hong,
  Norte, Slater, Shang, Krause, Anant, Aspelmeyer, and
  Gr{\"o}blacher}}]{Riedinger_nonclass_mechlight_nature}
\bibinfo{author}{\bibfnamefont{R.}~\bibnamefont{Riedinger}},
  \bibinfo{author}{\bibfnamefont{S.}~\bibnamefont{Hong}},
  \bibinfo{author}{\bibfnamefont{R.~A.} \bibnamefont{Norte}},
  \bibinfo{author}{\bibfnamefont{J.~A.} \bibnamefont{Slater}},
  \bibinfo{author}{\bibfnamefont{J.}~\bibnamefont{Shang}},
  \bibinfo{author}{\bibfnamefont{A.~G.} \bibnamefont{Krause}},
  \bibinfo{author}{\bibfnamefont{V.}~\bibnamefont{Anant}},
  \bibinfo{author}{\bibfnamefont{M.}~\bibnamefont{Aspelmeyer}},
  \bibnamefont{and}
  \bibinfo{author}{\bibfnamefont{S.}~\bibnamefont{Gr{\"o}blacher}},
  \bibinfo{journal}{Nature} \textbf{\bibinfo{volume}{530}},
  \bibinfo{pages}{313} (\bibinfo{year}{2016}).

\bibitem[{\citenamefont{Lecocq et~al.}(2015)\citenamefont{Lecocq, Clark,
  Simmonds, Aumentado, and Teufel}}]{Lecocq_QNDnonclassical_PRX}
\bibinfo{author}{\bibfnamefont{F.}~\bibnamefont{Lecocq}},
  \bibinfo{author}{\bibfnamefont{J.~B.} \bibnamefont{Clark}},
  \bibinfo{author}{\bibfnamefont{R.~W.} \bibnamefont{Simmonds}},
  \bibinfo{author}{\bibfnamefont{J.}~\bibnamefont{Aumentado}},
  \bibnamefont{and} \bibinfo{author}{\bibfnamefont{J.~D.}
  \bibnamefont{Teufel}}, \bibinfo{journal}{Phys. Rev. X}
  \textbf{\bibinfo{volume}{5}}, \bibinfo{pages}{041037} (\bibinfo{year}{2015}).

\bibitem[{\citenamefont{Lei et~al.}(2016)\citenamefont{Lei, Weinstein, Suh,
  Wollman, Kronwald, Marquardt, Clerk, and Schwab}}]{Lei_2016_Nondemolition}
\bibinfo{author}{\bibfnamefont{C.~U.} \bibnamefont{Lei}},
  \bibinfo{author}{\bibfnamefont{A.~J.} \bibnamefont{Weinstein}},
  \bibinfo{author}{\bibfnamefont{J.}~\bibnamefont{Suh}},
  \bibinfo{author}{\bibfnamefont{E.~E.} \bibnamefont{Wollman}},
  \bibinfo{author}{\bibfnamefont{A.}~\bibnamefont{Kronwald}},
  \bibinfo{author}{\bibfnamefont{F.}~\bibnamefont{Marquardt}},
  \bibinfo{author}{\bibfnamefont{A.~A.} \bibnamefont{Clerk}}, \bibnamefont{and}
  \bibinfo{author}{\bibfnamefont{K.~C.} \bibnamefont{Schwab}},
  \bibinfo{journal}{Phys. Rev. Lett.} \textbf{\bibinfo{volume}{117}},
  \bibinfo{pages}{100801} (\bibinfo{year}{2016}).

\bibitem[{\citenamefont{Vanner et~al.}(2015)\citenamefont{Vanner, Pikovski, and
  Kim}}]{vanner:statereconstr}
\bibinfo{author}{\bibfnamefont{M.~R.} \bibnamefont{Vanner}},
  \bibinfo{author}{\bibfnamefont{I.}~\bibnamefont{Pikovski}}, \bibnamefont{and}
  \bibinfo{author}{\bibfnamefont{M.~S.} \bibnamefont{Kim}},
  \bibinfo{journal}{Annalen der Physik} \textbf{\bibinfo{volume}{527}},
  \bibinfo{pages}{15} (\bibinfo{year}{2015}).

\bibitem[{\citenamefont{Woolley et~al.}(2010)\citenamefont{Woolley, Doherty,
  and Milburn}}]{Woolley_2010_quantum}
\bibinfo{author}{\bibfnamefont{M.~J.} \bibnamefont{Woolley}},
  \bibinfo{author}{\bibfnamefont{A.~C.} \bibnamefont{Doherty}},
  \bibnamefont{and} \bibinfo{author}{\bibfnamefont{G.~J.}
  \bibnamefont{Milburn}}, \bibinfo{journal}{Phys. Rev. B}
  \textbf{\bibinfo{volume}{82}}, \bibinfo{pages}{094511}
  (\bibinfo{year}{2010}).

\bibitem[{\citenamefont{Gangat et~al.}(2011)\citenamefont{Gangat, Stace, and
  Milburn}}]{Gangat_2011_number}
\bibinfo{author}{\bibfnamefont{A.~A.} \bibnamefont{Gangat}},
  \bibinfo{author}{\bibfnamefont{T.~M.} \bibnamefont{Stace}}, \bibnamefont{and}
  \bibinfo{author}{\bibfnamefont{G.~J.} \bibnamefont{Milburn}},
  \bibinfo{journal}{New Journal of Physics} \textbf{\bibinfo{volume}{13}},
  \bibinfo{pages}{043024} (\bibinfo{year}{2011}).

\bibitem[{\citenamefont{Ludwig et~al.}(2012)\citenamefont{Ludwig,
  Safavi-Naeini, Painter, and Marquardt}}]{Ludwig_2012_Quantum}
\bibinfo{author}{\bibfnamefont{Max}~\bibnamefont{Ludwig}},
  \bibinfo{author}{\bibfnamefont{Amir~H.} \bibnamefont{Safavi-Naeini}},
  \bibinfo{author}{\bibfnamefont{Oskar}~\bibnamefont{Painter}}, \bibnamefont{and}
  \bibinfo{author}{\bibfnamefont{Florian}~\bibnamefont{Marquardt}},
  \bibinfo{journal}{Phys. Rev. Lett.} \textbf{\bibinfo{volume}{109}}
  (\bibinfo{year}{2012}).

\bibitem[{\citenamefont{Yin et~al.}(2013)\citenamefont{Yin, Li, Zhang, and
  Duan}}]{Yin_2013_quantum}
\bibinfo{author}{\bibfnamefont{Z.Q.} \bibnamefont{Yin}},
  \bibinfo{author}{\bibfnamefont{T.}~\bibnamefont{Li}},
  \bibinfo{author}{\bibfnamefont{X.}~\bibnamefont{Zhang}}, \bibnamefont{and}
  \bibinfo{author}{\bibfnamefont{L.~M.} \bibnamefont{Duan}},
  \bibinfo{journal}{Phys. Rev. A} \textbf{\bibinfo{volume}{88}},
  \bibinfo{pages}{033614} (\bibinfo{year}{2013}).

\bibitem[{\citenamefont{Kim et~al.}(2016)\citenamefont{Kim, Hauer, Doolin,
  Souris, and Davis}}]{kim_quantlim_torque_NComm}
\bibinfo{author}{\bibfnamefont{P.~H.} \bibnamefont{Kim}},
  \bibinfo{author}{\bibfnamefont{B.~D.} \bibnamefont{Hauer}},
  \bibinfo{author}{\bibfnamefont{C.}~\bibnamefont{Doolin}},
  \bibinfo{author}{\bibfnamefont{F.}~\bibnamefont{Souris}}, \bibnamefont{and}
  \bibinfo{author}{\bibfnamefont{J.~P.} \bibnamefont{Davis}},
  \bibinfo{journal}{Nature Communications} \textbf{\bibinfo{volume}{7}},
  \bibinfo{pages}{13165 EP } (\bibinfo{year}{2016}).

\bibitem[{\citenamefont{Kim et~al.}(2013)\citenamefont{Kim, Doolin, Hauer,
  MacDonald, Freeman, Barclay, and Davis}}]{kim_torsional_optmech}
\bibinfo{author}{\bibfnamefont{P.~H.} \bibnamefont{Kim}},
  \bibinfo{author}{\bibfnamefont{C.}~\bibnamefont{Doolin}},
  \bibinfo{author}{\bibfnamefont{B.~D.} \bibnamefont{Hauer}},
  \bibinfo{author}{\bibfnamefont{A.~J.~R.} \bibnamefont{MacDonald}},
  \bibinfo{author}{\bibfnamefont{M.~R.} \bibnamefont{Freeman}},
  \bibinfo{author}{\bibfnamefont{P.~E.} \bibnamefont{Barclay}},
  \bibnamefont{and} \bibinfo{author}{\bibfnamefont{J.~P.} \bibnamefont{Davis}},
  \bibinfo{journal}{Applied Physics Letters} \textbf{\bibinfo{volume}{102}},
  \bibinfo{eid}{053102} (\bibinfo{year}{2013}).

\bibitem[{\citenamefont{Gallagher}(1994)}]{book:gallagher}
\bibinfo{author}{\bibfnamefont{T.~F.} \bibnamefont{Gallagher}},
  \emph{\bibinfo{title}{Rydberg Atoms}} (\bibinfo{publisher}{Cambridge
  University Press}, \bibinfo{year}{1994}).

\bibitem[{\citenamefont{Hogan et~al.}(2012{\natexlab{a}})\citenamefont{Hogan,
  Allmendinger, Sa\ss{}mannshausen, Schmutz, and Merkt}}]{Hogan_2011_Electrode}
\bibinfo{author}{\bibfnamefont{S.~D.} \bibnamefont{Hogan}},
  \bibinfo{author}{\bibfnamefont{P.}~\bibnamefont{Allmendinger}},
  \bibinfo{author}{\bibfnamefont{H.}~\bibnamefont{Sa\ss{}mannshausen}},
  \bibinfo{author}{\bibfnamefont{H.}~\bibnamefont{Schmutz}}, \bibnamefont{and}
  \bibinfo{author}{\bibfnamefont{F.}~\bibnamefont{Merkt}},
  \bibinfo{journal}{Phys. Rev. Lett.} \textbf{\bibinfo{volume}{108}},
  \bibinfo{pages}{063008} (\bibinfo{year}{2012}{\natexlab{a}}).

\bibitem[{\citenamefont{Lancuba and Hogan}(2013)}]{Lancuba_2013_Rydberg}
\bibinfo{author}{\bibfnamefont{P.}~\bibnamefont{Lancuba}} \bibnamefont{and}
  \bibinfo{author}{\bibfnamefont{S.~D.} \bibnamefont{Hogan}},
  \bibinfo{journal}{Phys. Rev. A} \textbf{\bibinfo{volume}{88}},
  \bibinfo{pages}{043427} (\bibinfo{year}{2013}).

\bibitem[{\citenamefont{Lancuba and Hogan}(2014)}]{Lancuba_2014_line}
\bibinfo{author}{\bibfnamefont{P.}~\bibnamefont{Lancuba}} \bibnamefont{and}
  \bibinfo{author}{\bibfnamefont{S.~D.} \bibnamefont{Hogan}},
  \bibinfo{journal}{Phys. Rev. A} \textbf{\bibinfo{volume}{90}},
  \bibinfo{pages}{053420} (\bibinfo{year}{2014}).

\bibitem[{\citenamefont{Carter et~al.}(2012)\citenamefont{Carter, Cherry, and
  Martin}}]{Carter_2012_field}
\bibinfo{author}{\bibfnamefont{J.~D.} \bibnamefont{Carter}},
  \bibinfo{author}{\bibfnamefont{O.}~\bibnamefont{Cherry}}, \bibnamefont{and}
  \bibinfo{author}{\bibfnamefont{J.~D.~D.} \bibnamefont{Martin}},
  \bibinfo{journal}{Phys. Rev. A} \textbf{\bibinfo{volume}{86}},
  \bibinfo{pages}{053401} (\bibinfo{year}{2012}).

\bibitem[{\citenamefont{Carter and Martin}(2013)}]{Carter_2013_manipulation}
\bibinfo{author}{\bibfnamefont{J.~D.} \bibnamefont{Carter}} \bibnamefont{and}
  \bibinfo{author}{\bibfnamefont{J.~D.~D.} \bibnamefont{Martin}},
  \bibinfo{journal}{Phys. Rev. A} \textbf{\bibinfo{volume}{88}},
  \bibinfo{pages}{043429} (\bibinfo{year}{2013}).

\bibitem[{\citenamefont{Thiele et~al.}(2014)\citenamefont{Thiele, Filipp,
  Agner, Schmutz, Deiglmayr, Stammeier, Allmendinger, Merkt, and
  Wallraff}}]{Thiele_2014_Rydberg}
\bibinfo{author}{\bibfnamefont{T.}~\bibnamefont{Thiele}},
  \bibinfo{author}{\bibfnamefont{S.}~\bibnamefont{Filipp}},
  \bibinfo{author}{\bibfnamefont{J.~A.} \bibnamefont{Agner}},
  \bibinfo{author}{\bibfnamefont{H.}~\bibnamefont{Schmutz}},
  \bibinfo{author}{\bibfnamefont{J.}~\bibnamefont{Deiglmayr}},
  \bibinfo{author}{\bibfnamefont{M.}~\bibnamefont{Stammeier}},
  \bibinfo{author}{\bibfnamefont{P.}~\bibnamefont{Allmendinger}},
  \bibinfo{author}{\bibfnamefont{F.}~\bibnamefont{Merkt}}, \bibnamefont{and}
  \bibinfo{author}{\bibfnamefont{A.}~\bibnamefont{Wallraff}},
  \bibinfo{journal}{Phys. Rev. A} \textbf{\bibinfo{volume}{90}},
  \bibinfo{pages}{013414} (\bibinfo{year}{2014}).

\bibitem[{\citenamefont{Hermann-Avigliano
  et~al.}(2014)\citenamefont{Hermann-Avigliano, Teixeira, Nguyen,
  Cantat-Moltrecht, Nogues, Dotsenko, Gleyzes, Raimond, Haroche, and
  Brune}}]{Avigliano_2014_coherence}
\bibinfo{author}{\bibfnamefont{C.}~\bibnamefont{Hermann-Avigliano}},
  \bibinfo{author}{\bibfnamefont{R.~C.} \bibnamefont{Teixeira}},
  \bibinfo{author}{\bibfnamefont{T.~L.} \bibnamefont{Nguyen}},
  \bibinfo{author}{\bibfnamefont{T.}~\bibnamefont{Cantat-Moltrecht}},
  \bibinfo{author}{\bibfnamefont{G.}~\bibnamefont{Nogues}},
  \bibinfo{author}{\bibfnamefont{I.}~\bibnamefont{Dotsenko}},
  \bibinfo{author}{\bibfnamefont{S.}~\bibnamefont{Gleyzes}},
  \bibinfo{author}{\bibfnamefont{J.~M.} \bibnamefont{Raimond}},
  \bibinfo{author}{\bibfnamefont{S.}~\bibnamefont{Haroche}}, \bibnamefont{and}
  \bibinfo{author}{\bibfnamefont{M.}~\bibnamefont{Brune}},
  \bibinfo{journal}{Phys. Rev. A} \textbf{\bibinfo{volume}{90}},
  \bibinfo{pages}{040502} (\bibinfo{year}{2014}).

\bibitem[{\citenamefont{Keil et~al.}(2016)\citenamefont{Keil, Amit, Zhou,
  Groswasser, Japha, and Folman}}]{keil:atomchip:review}
\bibinfo{author}{\bibfnamefont{M.}~\bibnamefont{Keil}},
  \bibinfo{author}{\bibfnamefont{O.}~\bibnamefont{Amit}},
  \bibinfo{author}{\bibfnamefont{S.}~\bibnamefont{Zhou}},
  \bibinfo{author}{\bibfnamefont{D.}~\bibnamefont{Groswasser}},
  \bibinfo{author}{\bibfnamefont{Y.}~\bibnamefont{Japha}}, \bibnamefont{and}
  \bibinfo{author}{\bibfnamefont{R.}~\bibnamefont{Folman}},
  \bibinfo{journal}{Journal of Modern Optics} \textbf{\bibinfo{volume}{63}},
  \bibinfo{pages}{1840} (\bibinfo{year}{2016}).

\bibitem[{\citenamefont{Hogan}(2016)}]{Hogan_2016_Stark}
\bibinfo{author}{\bibfnamefont{S.~D.} \bibnamefont{Hogan}},
  \bibinfo{journal}{{EPJ} Techniques and Instrumentation}
  \textbf{\bibinfo{volume}{3}} (\bibinfo{year}{2016}).

\bibitem[{\citenamefont{Brune et~al.}(1990)\citenamefont{Brune, Haroche,
  Lefevre, Raimond, and Zagury}}]{BrHaLe90_976}
\bibinfo{author}{\bibfnamefont{M.}~\bibnamefont{Brune}},
  \bibinfo{author}{\bibfnamefont{S.}~\bibnamefont{Haroche}},
  \bibinfo{author}{\bibfnamefont{V.}~\bibnamefont{Lefevre}},
  \bibinfo{author}{\bibfnamefont{J.M.}~\bibnamefont{Raimond}}, \bibnamefont{and}
  \bibinfo{author}{\bibfnamefont{N.}~\bibnamefont{Zagury}},
  \bibinfo{journal}{Phys. Rev. Lett.} \textbf{\bibinfo{volume}{{65}}},
  \bibinfo{pages}{976} (\bibinfo{year}{1990}).

\bibitem[{\citenamefont{Brune et~al.}(2008)\citenamefont{Brune, Bernu, Guerlin,
  Del{\'e}glise, Sayrin, Gleyzes, Kuhr, Dotsenko, Raimond, and
  Haroche}}]{brune:processtomog}
\bibinfo{author}{\bibfnamefont{M.}~\bibnamefont{Brune}},
  \bibinfo{author}{\bibfnamefont{J.}~\bibnamefont{Bernu}},
  \bibinfo{author}{\bibfnamefont{C.}~\bibnamefont{Guerlin}},
  \bibinfo{author}{\bibfnamefont{S.}~\bibnamefont{Del{\'e}glise}},
  \bibinfo{author}{\bibfnamefont{C.}~\bibnamefont{Sayrin}},
  \bibinfo{author}{\bibfnamefont{S.}~\bibnamefont{Gleyzes}},
  \bibinfo{author}{\bibfnamefont{S.}~\bibnamefont{Kuhr}},
  \bibinfo{author}{\bibfnamefont{I.}~\bibnamefont{Dotsenko}},
  \bibinfo{author}{\bibfnamefont{J.~M.} \bibnamefont{Raimond}},
  \bibnamefont{and} \bibinfo{author}{\bibfnamefont{S.}~\bibnamefont{Haroche}},
  \bibinfo{journal}{Phys. Rev. Lett.} \textbf{\bibinfo{volume}{{101}}},
  \bibinfo{pages}{240402} (\bibinfo{year}{2008}).

\bibitem[{\citenamefont{Brune et~al.}(1992)\citenamefont{Brune, Haroche,
  Raimond, Davidovich, and Zagury}}]{BrHaRa92_5193}
\bibinfo{author}{\bibfnamefont{M.}~\bibnamefont{Brune}},
  \bibinfo{author}{\bibfnamefont{S.}~\bibnamefont{Haroche}},
  \bibinfo{author}{\bibfnamefont{J.~M.}~\bibnamefont{Raimond}},
  \bibinfo{author}{\bibfnamefont{L.}~\bibnamefont{Davidovich}},
  \bibnamefont{and} \bibinfo{author}{\bibfnamefont{N.}~\bibnamefont{Zagury}},
  \bibinfo{journal}{Phys. Rev. A} \textbf{\bibinfo{volume}{{45}}},
  \bibinfo{pages}{5193} (\bibinfo{year}{1992}).

\bibitem[{\citenamefont{Gleyzes et~al.}(2007)\citenamefont{Gleyzes, Kuhr,
  Guerlin, Bernu, Del{\'e}glise, Hoff, Brune, Raimond, and
  Haroche}}]{GlKuGu07_297}
\bibinfo{author}{\bibfnamefont{S.}~\bibnamefont{Gleyzes}},
  \bibinfo{author}{\bibfnamefont{S.}~\bibnamefont{Kuhr}},
  \bibinfo{author}{\bibfnamefont{C.}~\bibnamefont{Guerlin}},
  \bibinfo{author}{\bibfnamefont{J.}~\bibnamefont{Bernu}},
  \bibinfo{author}{\bibfnamefont{S.}~\bibnamefont{Del{\'e}glise}},
  \bibinfo{author}{\bibfnamefont{U.~B.} \bibnamefont{Hoff}},
  \bibinfo{author}{\bibfnamefont{M.}~\bibnamefont{Brune}},
  \bibinfo{author}{\bibfnamefont{J.-M.} \bibnamefont{Raimond}},
  \bibnamefont{and} \bibinfo{author}{\bibfnamefont{S.}~\bibnamefont{Haroche}},
  \bibinfo{journal}{Nature} \textbf{\bibinfo{volume}{{446}}},
  \bibinfo{pages}{297} (\bibinfo{year}{2007}).

\bibitem[{\citenamefont{Del{\'e}glise et~al.}(2008)\citenamefont{Del{\'e}glise,
  Dotsenko, Sayrin, Bernu, Raimond, and Haroche}}]{deleglise:reconstruction}
\bibinfo{author}{\bibfnamefont{S.}~\bibnamefont{Del{\'e}glise}},
  \bibinfo{author}{\bibfnamefont{I.}~\bibnamefont{Dotsenko}},
  \bibinfo{author}{\bibfnamefont{C.}~\bibnamefont{Sayrin}},
  \bibinfo{author}{\bibfnamefont{J.}~\bibnamefont{Bernu}},
  \bibinfo{author}{\bibfnamefont{J.-M.} \bibnamefont{Raimond}},
  \bibnamefont{and} \bibinfo{author}{\bibfnamefont{S.}~\bibnamefont{Haroche}},
  \bibinfo{journal}{Nature} \textbf{\bibinfo{volume}{{455}}},
  \bibinfo{pages}{510} (\bibinfo{year}{2008}).

\bibitem[{\citenamefont{Singh and Meystre}(2013)}]{singh:wignerntomog}
\bibinfo{author}{\bibfnamefont{S.}~\bibnamefont{Singh}} \bibnamefont{and}
  \bibinfo{author}{\bibfnamefont{P.}~\bibnamefont{Meystre}},
  \bibinfo{journal}{Phys. Rev. A} \textbf{\bibinfo{volume}{81}},
  \bibinfo{pages}{041804(R)} (\bibinfo{year}{2010}).

\bibitem[{\citenamefont{Singh et~al.}(2008)\citenamefont{Singh, Bhattacharya,
  Dutta, and Meystre}}]{singh:cantilevermol}
\bibinfo{author}{\bibfnamefont{S.}~\bibnamefont{Singh}},
  \bibinfo{author}{\bibfnamefont{M.}~\bibnamefont{Bhattacharya}},
  \bibinfo{author}{\bibfnamefont{O.}~\bibnamefont{Dutta}}, \bibnamefont{and}
  \bibinfo{author}{\bibfnamefont{P.}~\bibnamefont{Meystre}},
  \bibinfo{journal}{Phys. Rev. Lett.} \textbf{\bibinfo{volume}{101}},
  \bibinfo{pages}{263603} (\bibinfo{year}{2008}).

\bibitem[{\citenamefont{Bariani et~al.}(2014)\citenamefont{Bariani, Otterbach,
  Tan, and Meystre}}]{bariani:controlwsingleatom}
\bibinfo{author}{\bibfnamefont{F.}~\bibnamefont{Bariani}},
  \bibinfo{author}{\bibfnamefont{J.}~\bibnamefont{Otterbach}},
  \bibinfo{author}{\bibfnamefont{H.}~\bibnamefont{Tan}}, \bibnamefont{and}
  \bibinfo{author}{\bibfnamefont{P.}~\bibnamefont{Meystre}},
  \bibinfo{journal}{Phys. Rev. A} \textbf{\bibinfo{volume}{89}},
  \bibinfo{pages}{011801(R)} (\bibinfo{year}{2014}).

\bibitem[{\citenamefont{Stevenson et~al.}(2016)\citenamefont{Stevenson,
  Min\'a\ifmmode~\check{r}\else \v{r}\fi{}, Hofferberth, and
  Lesanovsky}}]{stevenson:chargedosci:rydberg}
\bibinfo{author}{\bibfnamefont{R.}~\bibnamefont{Stevenson}},
  \bibinfo{author}{\bibfnamefont{J.}
  \bibnamefont{Min\'a\ifmmode~\check{r}\else \v{r}\fi{}}},
  \bibinfo{author}{\bibfnamefont{S.}~\bibnamefont{Hofferberth}},
  \bibnamefont{and}
  \bibinfo{author}{\bibfnamefont{I.}~\bibnamefont{Lesanovsky}},
  \bibinfo{journal}{Phys. Rev. A} \textbf{\bibinfo{volume}{94}},
  \bibinfo{pages}{043813} (\bibinfo{year}{2016}).

\bibitem[{\citenamefont{Meyer et~al.}(2005)\citenamefont{Meyer, Paillet, and
  Roth}}]{MePaRo05_1539}
\bibinfo{author}{\bibfnamefont{J.}~\bibnamefont{Meyer}},
  \bibinfo{author}{\bibfnamefont{M.}~\bibnamefont{Paillet}}, \bibnamefont{and}
  \bibinfo{author}{\bibfnamefont{S.}~\bibnamefont{Roth}},
  \bibinfo{journal}{Science} \textbf{\bibinfo{volume}{{309}}},
  \bibinfo{pages}{1539} (\bibinfo{year}{2005}).

\bibitem[{\citenamefont{Ganzhorn et~al.}(2013)\citenamefont{Ganzhorn,
  Klyatskaya, Ruben, and Wernsdorfer}}]{Ganzhorn_2013_spin}
\bibinfo{author}{\bibfnamefont{M.}~\bibnamefont{Ganzhorn}},
  \bibinfo{author}{\bibfnamefont{S.}~\bibnamefont{Klyatskaya}},
  \bibinfo{author}{\bibfnamefont{M.}~\bibnamefont{Ruben}}, \bibnamefont{and}
  \bibinfo{author}{\bibfnamefont{W.}~\bibnamefont{Wernsdorfer}},
  \bibinfo{journal}{Nature Nanotech} \textbf{\bibinfo{volume}{8}},
  \bibinfo{pages}{165} (\bibinfo{year}{2013}).

\bibitem[{\citenamefont{Basun et~al.}(2011)\citenamefont{Basun, Cook,
  Reshetnyak, Glushchenko, and Evans}}]{Basun_2011_moment}
\bibinfo{author}{\bibfnamefont{S.~A.} \bibnamefont{Basun}},
  \bibinfo{author}{\bibfnamefont{G.}~\bibnamefont{Cook}},
  \bibinfo{author}{\bibfnamefont{V.~Y.} \bibnamefont{Reshetnyak}},
  \bibinfo{author}{\bibfnamefont{A.~V.} \bibnamefont{Glushchenko}},
  \bibnamefont{and} \bibinfo{author}{\bibfnamefont{D.~R.} \bibnamefont{Evans}},
  \bibinfo{journal}{Phys. Rev. B} \textbf{\bibinfo{volume}{84}},
  \bibinfo{pages}{024105} (\bibinfo{year}{2011}).

\bibitem[{\citenamefont{Hogan et~al.}(2012{\natexlab{b}})\citenamefont{Hogan,
  Agner, Merkt, Thiele, Filipp, and Wallraff}}]{Hogan_2012_Rydberg}
\bibinfo{author}{\bibfnamefont{S.~D.} \bibnamefont{Hogan}},
  \bibinfo{author}{\bibfnamefont{J.~A.} \bibnamefont{Agner}},
  \bibinfo{author}{\bibfnamefont{F.}~\bibnamefont{Merkt}},
  \bibinfo{author}{\bibfnamefont{T.}~\bibnamefont{Thiele}},
  \bibinfo{author}{\bibfnamefont{S.}~\bibnamefont{Filipp}}, \bibnamefont{and}
  \bibinfo{author}{\bibfnamefont{A.}~\bibnamefont{Wallraff}},
  \bibinfo{journal}{Phys. Rev. Lett.} \textbf{\bibinfo{volume}{108}}
  (\bibinfo{year}{2012}{\natexlab{b}}).

\bibitem[{\citenamefont{Guerlin et~al.}(2007)\citenamefont{Guerlin, Bernu,
  Del{\'{e}}glise, Sayrin, Gleyzes, Kuhr, Brune, Raimond, and
  Haroche}}]{Guerlin_2007_field}
\bibinfo{author}{\bibfnamefont{C.}~\bibnamefont{Guerlin}},
  \bibinfo{author}{\bibfnamefont{J.}~\bibnamefont{Bernu}},
  \bibinfo{author}{\bibfnamefont{S.}~\bibnamefont{Del{\'{e}}glise}},
  \bibinfo{author}{\bibfnamefont{C.}~\bibnamefont{Sayrin}},
  \bibinfo{author}{\bibfnamefont{S.}~\bibnamefont{Gleyzes}},
  \bibinfo{author}{\bibfnamefont{S.}~\bibnamefont{Kuhr}},
  \bibinfo{author}{\bibfnamefont{M.}~\bibnamefont{Brune}},
  \bibinfo{author}{\bibfnamefont{J.-M.} \bibnamefont{Raimond}},
  \bibnamefont{and} \bibinfo{author}{\bibfnamefont{S.}~\bibnamefont{Haroche}},
  \bibinfo{journal}{Nature} \textbf{\bibinfo{volume}{448}},
  \bibinfo{pages}{889} (\bibinfo{year}{2007}).

\bibitem[{\citenamefont{Beck et~al.}(2016)\citenamefont{Beck, Isaacs, Booth,
  Pritchard, Saffman, and McDermott}}]{Beck_2016_coplanar}
\bibinfo{author}{\bibfnamefont{M.~A.} \bibnamefont{Beck}},
  \bibinfo{author}{\bibfnamefont{J.~A.} \bibnamefont{Isaacs}},
  \bibinfo{author}{\bibfnamefont{D.}~\bibnamefont{Booth}},
  \bibinfo{author}{\bibfnamefont{J.~D.} \bibnamefont{Pritchard}},
  \bibinfo{author}{\bibfnamefont{M.}~\bibnamefont{Saffman}}, \bibnamefont{and}
  \bibinfo{author}{\bibfnamefont{R.}~\bibnamefont{McDermott}},
  \bibinfo{journal}{Appl. Phys. Lett.} \textbf{\bibinfo{volume}{109}},
  \bibinfo{pages}{092602} (\bibinfo{year}{2016}).

\bibitem[{\citenamefont{Thiele et~al.}(2015)\citenamefont{Thiele, Deiglmayr,
  Stammeier, Agner, Schmutz, Merkt, and Wallraff}}]{Thiele_2015_electric}
\bibinfo{author}{\bibfnamefont{T.}~\bibnamefont{Thiele}},
  \bibinfo{author}{\bibfnamefont{J.}~\bibnamefont{Deiglmayr}},
  \bibinfo{author}{\bibfnamefont{M.}~\bibnamefont{Stammeier}},
  \bibinfo{author}{\bibfnamefont{J.-A.} \bibnamefont{Agner}},
  \bibinfo{author}{\bibfnamefont{H.}~\bibnamefont{Schmutz}},
  \bibinfo{author}{\bibfnamefont{F.}~\bibnamefont{Merkt}}, \bibnamefont{and}
  \bibinfo{author}{\bibfnamefont{A.}~\bibnamefont{Wallraff}},
  \bibinfo{journal}{Phys. Rev. A} \textbf{\bibinfo{volume}{92}},
  \bibinfo{pages}{063425} (\bibinfo{year}{2015}).

\bibitem[{\citenamefont{de~Melo and Vianna}(2015)}]{Melo_2015_shift}
\bibinfo{author}{\bibfnamefont{N.~R.} \bibnamefont{de~Melo}} \bibnamefont{and}
  \bibinfo{author}{\bibfnamefont{S.~S.} \bibnamefont{Vianna}},
  \bibinfo{journal}{Phys. Rev. A} \textbf{\bibinfo{volume}{92}},
  \bibinfo{pages}{053830} (\bibinfo{year}{2015}).

\bibitem[{sup()}]{sup:inf}
\bibinfo{note}{See Supplemental Material for details on Hamiltonian, oscillator driving, design criteria and
  parameters.}

\bibitem[{\citenamefont{Haroche and Raimond}(2006)}]{Haroche_2006_the}
\bibinfo{author}{\bibfnamefont{S.}~\bibnamefont{Haroche}} \bibnamefont{and}
  \bibinfo{author}{\bibfnamefont{J.-M.} \bibnamefont{Raimond}},
  \emph{\bibinfo{title}{Exploring the quantum: atoms, cavities, and photons}}
  (\bibinfo{publisher}{Oxford university press}, \bibinfo{year}{2006}).

\end{thebibliography}

\begin{thebibliography}{8}
\expandafter\ifx\csname natexlab\endcsname\relax\def\natexlab#1{#1}\fi
\expandafter\ifx\csname bibnamefont\endcsname\relax
  \def\bibnamefont#1{#1}\fi
\expandafter\ifx\csname bibfnamefont\endcsname\relax
  \def\bibfnamefont#1{#1}\fi
\expandafter\ifx\csname citenamefont\endcsname\relax
  \def\citenamefont#1{#1}\fi
\expandafter\ifx\csname url\endcsname\relax
  \def\url#1{\texttt{#1}}\fi
\expandafter\ifx\csname urlprefix\endcsname\relax\def\urlprefix{URL }\fi
\providecommand{\bibinfo}[2]{#2}
\providecommand{\eprint}[2][]{\url{#2}}

\bibitem[{\citenamefont{Meyer et~al.}(2005)\citenamefont{Meyer, Paillet, and
  Roth}}]{MePaRo05_1539}
\bibinfo{author}{\bibfnamefont{J.}~\bibnamefont{Meyer}},
  \bibinfo{author}{\bibfnamefont{M.}~\bibnamefont{Paillet}}, \bibnamefont{and}
  \bibinfo{author}{\bibfnamefont{S.}~\bibnamefont{Roth}},
  \bibinfo{journal}{Science} \textbf{\bibinfo{volume}{{309}}},
  \bibinfo{pages}{1539} (\bibinfo{year}{2005}).

\bibitem[{foo()}]{footnote:transitiondipolefixed}
\bibinfo{note}{We can for example define a quantisation axis $||$ $z$ through
  Rydberg excitation laser polarisation and work with $m_j=1/2$ states only.}

\bibitem[{\citenamefont{Brune et~al.}(1990)\citenamefont{Brune, Haroche,
  Lefevre, Raimond, and Zagury}}]{Brune_1990_nondemolition}
\bibinfo{author}{\bibfnamefont{M.}~\bibnamefont{Brune}},
  \bibinfo{author}{\bibfnamefont{S.}~\bibnamefont{Haroche}},
  \bibinfo{author}{\bibfnamefont{V.}~\bibnamefont{Lefevre}},
  \bibinfo{author}{\bibfnamefont{J.~M.} \bibnamefont{Raimond}},
  \bibnamefont{and} \bibinfo{author}{\bibfnamefont{N.}~\bibnamefont{Zagury}},
  \bibinfo{journal}{Phys. Rev. Lett.} \textbf{\bibinfo{volume}{65}},
  \bibinfo{pages}{976} (\bibinfo{year}{1990}).

\bibitem[{\citenamefont{Gardiner and Zoller}(2004)}]{Gardiner_2004_noise}
\bibinfo{author}{\bibfnamefont{C.}~\bibnamefont{Gardiner}} \bibnamefont{and}
  \bibinfo{author}{\bibfnamefont{P.}~\bibnamefont{Zoller}},
  \emph{\bibinfo{title}{Quantum noise: a handbook of Markovian and
  non-Markovian quantum stochastic methods with applications to quantum
  optics}}, vol.~\bibinfo{volume}{56} (\bibinfo{publisher}{Springer Science \&
  Business Media}, \bibinfo{year}{2004}).

\bibitem[{\citenamefont{Beterov et~al.}(2009)\citenamefont{Beterov, Ryabtsev,
  Tretyakov, and Entin}}]{Beterov_2009_calculations}
\bibinfo{author}{\bibfnamefont{I.~I.} \bibnamefont{Beterov}},
  \bibinfo{author}{\bibfnamefont{I.~I.} \bibnamefont{Ryabtsev}},
  \bibinfo{author}{\bibfnamefont{D.~B.} \bibnamefont{Tretyakov}},
  \bibnamefont{and} \bibinfo{author}{\bibfnamefont{V.~M.} \bibnamefont{Entin}},
  \bibinfo{journal}{Phys. Rev. A} \textbf{\bibinfo{volume}{79}},
  \bibinfo{pages}{052504} (\bibinfo{year}{2009}).

\bibitem[{\citenamefont{Carter and Martin}(2013)}]{Carter_2013_manipulation}
\bibinfo{author}{\bibfnamefont{J.~D.} \bibnamefont{Carter}} \bibnamefont{and}
  \bibinfo{author}{\bibfnamefont{J.~D.~D.} \bibnamefont{Martin}},
  \bibinfo{journal}{Phys. Rev. A} \textbf{\bibinfo{volume}{88}},
  \bibinfo{pages}{043429} (\bibinfo{year}{2013}).

\bibitem[{\citenamefont{Hermann-Avigliano
  et~al.}(2014)\citenamefont{Hermann-Avigliano, Teixeira, Nguyen,
  Cantat-Moltrecht, Nogues, Dotsenko, Gleyzes, Raimond, Haroche, and
  Brune}}]{Avigliano_2014_coherence}
\bibinfo{author}{\bibfnamefont{C.}~\bibnamefont{Hermann-Avigliano}},
  \bibinfo{author}{\bibfnamefont{R.~C.} \bibnamefont{Teixeira}},
  \bibinfo{author}{\bibfnamefont{T.~L.} \bibnamefont{Nguyen}},
  \bibinfo{author}{\bibfnamefont{T.}~\bibnamefont{Cantat-Moltrecht}},
  \bibinfo{author}{\bibfnamefont{G.}~\bibnamefont{Nogues}},
  \bibinfo{author}{\bibfnamefont{I.}~\bibnamefont{Dotsenko}},
  \bibinfo{author}{\bibfnamefont{S.}~\bibnamefont{Gleyzes}},
  \bibinfo{author}{\bibfnamefont{J.~M.} \bibnamefont{Raimond}},
  \bibinfo{author}{\bibfnamefont{S.}~\bibnamefont{Haroche}}, \bibnamefont{and}
  \bibinfo{author}{\bibfnamefont{M.}~\bibnamefont{Brune}},
  \bibinfo{journal}{Phys. Rev. A} \textbf{\bibinfo{volume}{90}},
  \bibinfo{pages}{040502} (\bibinfo{year}{2014}).

\bibitem[{\citenamefont{Wiseman and Milburn}(2009)}]{Wiseman_2009_measurement}
\bibinfo{author}{\bibfnamefont{H.~M.} \bibnamefont{Wiseman}} \bibnamefont{and}
  \bibinfo{author}{\bibfnamefont{G.~J.} \bibnamefont{Milburn}},
  \emph{\bibinfo{title}{Quantum measurement and control}}
  (\bibinfo{publisher}{Cambridge University Press}, \bibinfo{year}{2009}).

\end{thebibliography}
\end{document}